\documentclass[acmsmall,screen]{acmart}

\setcopyright{cc}
\setcctype{by-nd}
\acmJournal{PACMMOD}
\acmYear{2025} \acmVolume{3} \acmNumber{3 (SIGMOD)} \acmArticle{224} \acmMonth{6} \acmPrice{}\acmDOI{10.1145/3725361}

\received{October 2024}
\received[revised]{January 2025}
\received[accepted]{February 2025}

\usepackage[ruled,vlined]{algorithm2e}
\usepackage{multicol}
\usepackage{multirow}
\usepackage{tablefootnote}

\usepackage[super]{nth}

\usepackage{tikz}
\usepackage{xcolor}

\usepackage{tabularx}
\newcolumntype{M}[1]{>{\centering\arraybackslash}m{#1}} % centered
\newcolumntype{R}[1]{>{\raggedleft\arraybackslash}m{#1}} % right centered
\newcolumntype{L}[1]{>{\raggedright\arraybackslash}m{#1}} % right centered
\usepackage{booktabs}

\usepackage{arydshln}

\usepackage{wrapfig}
\usepackage{graphicx}
\usepackage{subcaption}
\usepackage{amsmath}

\usepackage{enumitem}
\setlist[itemize]{align=parleft,left=0pt..1.2em} % https://tex.stackexchange.com/questions/91124/itemize-removing-natural-indent

\begin{document}

\title{SwiftSpatial: Spatial Joins on Modern Hardware}

% %%
% %% The "author" command and its associated commands are used to define the authors and their affiliations.
\author{Wenqi Jiang}
\affiliation{%
  % \institution{Systems Group, Department of Computer Science, ETH Zurich}
  \institution{Systems Group, ETH Zurich}
  \city{Zurich}
  \country{Switzerland}
}
\email{wenqi.jiang@inf.ethz.ch}

\author{Oleh-Yevhen Khavrona}
\affiliation{%
  % \institution{Systems Group, Department of Computer Science, ETH Zurich}
  \institution{Systems Group, ETH Zurich}
  \city{Zurich}
  \country{Switzerland}
}
\email{yevhen.khavrona@inf.ethz.ch}

\author{Martin Parvanov}
\affiliation{%
  % \institution{Systems Group, Department of Computer Science, ETH Zurich}
  \institution{Systems Group, ETH Zurich}
  \city{Zurich}
  \country{Switzerland}
}
\email{martin.ivov.parvanov@gmail.com}

\author{Gustavo Alonso}
\affiliation{%
  % \institution{Systems Group, Department of Computer Science, ETH Zurich}
  \institution{Systems Group, ETH Zurich}
  \city{Zurich}
  \country{Switzerland}
}
\email{alonso@inf.ethz.ch}


\sloppy
\begin{abstract}

Spatial joins are among the most time-consuming spatial queries, remaining costly even in parallel and distributed systems.
In this paper, we explore hardware acceleration for spatial joins by proposing \textit{SwiftSpatial}, an FPGA-based accelerator that can be deployed in data centers and at the edge.
SwiftSpatial contains multiple high-performance join units with innovative hybrid parallelism, several efficient memory management units, and an extensible on-chip join scheduler that supports the popular R-tree synchronous traversal and partition-based spatial-merge (PBSM) algorithms.
Benchmarked against various CPU and GPU-based spatial data processing systems, SwiftSpatial demonstrates a latency reduction of up to 41.03$\times$ relative to the best-performing baseline, while requiring 6.16$\times$ less power. 
The performance and energy efficiency of SwiftSpatial demonstrate its potential to be used in a variety of configurations (e.g., as an accelerator, near storage, in-network) as well as on different devices (e.g., data centers where FPGAs are widely available or mobile devices, which also contain FPGAs for specialized processing).

% , an FPGA-based hardware accelerator for spatial join operations.
% in data centers and at the edge.

% We prototype SwiftSpatial on an FPGA and incorporate the R-tree synchronous traversal algorithm as the control flow. 

\end{abstract}

% %%
% %% The code below is generated by the tool at http://dl.acm.org/ccs.cfm.
% %% Please copy and paste the code instead of the example below.


% %%
% %% Keywords. The author(s) should pick words that accurately describe
% %% the work being presented. Separate the keywords with commas.
  
\maketitle

\section{Introduction}
\label{sec:intro}

Spatial joins --- where two sets of spatial objects are joined based on their spatial relationships --- are time-consuming spatial queries even if many parallel and distributed spatial data processing systems have been developed~\cite{pandey2018good, aji2013hadoop, eldawy2015spatial}. For instance, a 2013 study on spatial big data frameworks~\cite{aji2013hadoop} reported that joining two indexed datasets, totaling fewer than 100 million records, required 10 minutes (equivalent to 32 CPU hours) when executed on eight servers equipped with 192 cores and 1TB of memory in total. Such performance has become insufficient considering the increasing number of moving objects, such as autonomous vehicles and IoT devices~\cite{sowell2013experimental, vsidlauskas2014spatial}. 
After decades of extensive research on efficient indexes and algorithms for spatial joins, further performance gains have become increasingly difficult given a certain computational budget.
This prompts the question: \textit{is it possible to improve spatial join performance without demanding more resources such as CPU cores or memory capacity?}

Given the increasing computational demands of spatial joins and the slowdown in Moore's law, we approach spatial joins from a hardware acceleration perspective in this paper. Specifically, we choose FPGAs as the hardware platform for the designed accelerator due to their growing adoption in the cloud, as evidenced by deployments at Microsoft~\cite{firestone2018azure}, Amazon~\cite{aqua}, or Alibaba~\cite{li2019cloud}.

Spatial join algorithms generally follow the following paradigm.
During the preprocessing phase, both input datasets are decomposed into small tiles. For example, the datasets can be partitioned using a simple grid~\cite{patel1996partition, tsitsigkos2019parallel} or indexed by a tree~\cite{guttman1984r, sellis1987r+, nobari2013touch}.
After preprocessing the datasets, the join process begins, which comprises two primary components. 
Firstly, a control flow identifies tile pairs from both datasets that could potentially produce results. 
For example, when both datasets are partitioned using a grid, the control flow selects identical tiles from both datasets; for datasets indexed by R-trees, synchronous traversal~\cite{brinkhoff1993efficient} is adopted to navigate both trees to identify intersected tile pairs.
Secondly, for each candidate tile pair, all objects in the tiles are evaluated based on the predicate to produce results or intermediate outcomes. The repetitive nature of these operations suggests that they could be amenable to parallelization using parallel processing units and pipeline parallelism, both available on FPGAs.

Thus, we propose \textit{SwiftSpatial}, an in-memory FPGA-based accelerator designed for high-performance and energy-efficient spatial joins.  
SwiftSpatial supports two widely used spatial join algorithms: R-tree synchronous traversal~\cite{brinkhoff1993efficient} and partition-based spatial-merge (PBSM)~\cite{patel1996partition}, and can be extended to support other spatial join algorithms due to its modular design.
With R-tree synchronous traversal, SwiftSpatial can leverage the R-trees already maintained by various spatial data processing systems, such as PostGIS~\cite{postgis}, Oracle Spatial~\cite{kothuri2002quadtree}, Apache Sedona~\cite{sedona}, and SpatialSpark~\cite{spatialspark}.
The support for PBSM is motivated by its scalability on parallel architectures, as already shown on multi-core CPUs~\cite{tsitsigkos2019parallel}.

\textbf{The key idea of SwiftSpatial is to build efficient hardware primitives for spatial joins by instantiating multiple specialized join units.}
\textbf{(C1)} Each SwiftSpatial join unit is optimized to achieve minimal latency when joining two spatial object tiles. Its superior performance stems from the innovative architecture that employs hybrid parallelism : (a) operator parallelism for fast predicate evaluation involving multiple object boundary comparisons and (b) pipeline parallelism between input ingestion, join predicate evaluation, and result production for qualifying object pairs. The hybrid parallelism allows a join unit to process an object pair every clock cycle, leading to high and predictable tile-level join performance.
Multiple join units can be instantiated to parallelize joins across different tiles. 
\textbf{(C2)} Moreover, SwiftSpatial incorporates an on-chip scheduler to orchestrate these join units, passing control signals directly to the join units, thus supporting complex control flow with minimal performance overhead. 
\textbf{(C3)} In addition to the join and control units, SwiftSpatial features specialized memory management units to further improve join performance. By directly managing the physical address space, the memory management units (a) optimize memory access patterns through memory request batching and memory address rewriting, and (b) avoid the overhead associated with page table management and dynamic memory allocation prevalent in software systems.

We evaluate SwiftSpatial on real-world and synthetic datasets of various size scales and compare it against a variety of CPU and GPU-based spatial data processing systems, including PostGIS~\cite{postgis}, Apache Sedona~\cite{sedona}, SpatialSpark~\cite{spatialspark}, cuSpatial~\cite{cuspatial}, and our multi-threaded C++ implementations of R-tree synchronous traversal and PBSM. 
Overall, SwiftSpatial demonstrates outstanding performance and energy efficiency, achieving up to 41.03$\times$ latency reduction compared to the best-performing baseline system while consuming only 23.48W of power, 6.16$\times$ less than the baseline CPU. 
Moreover, the modular design of SwiftSpatial suggests it can not only be instantiated on data-center-grade FPGAs but also has the potential for deployment on embedded systems, showing its versatility and wide-ranging applicability.

The contributions of the paper are as follows:

 \begin{itemize}
     \item We design SwiftSpatial, a spatial join accelerator with:
     \begin{itemize}
     % \vspace{-0.5em}
        \item Spatial join units with innovative hybrid parallelism.
     % \vspace{-0.2em}
        \item Multiple efficient memory management units.
     % \vspace{-0.2em}
        \item An on-chip spatial join scheduler that coordinates join units and memory management units.
     \end{itemize}
     % \vspace{-0.5em}
     \item We implement SwiftSpatial on FPGA and support R-tree synchronous traversal and PBSM algorithms as the control flows.
     % \vspace{-0.5em}
     \item We compare SwiftSpatial with CPU and GPU-based spatial data processing systems on various datasets, showcasing its performance and energy efficiency.
 \end{itemize}

\section{Background and Motivation}
\label{sec:background}

% \vspace{-1em}
\subsection{Spatial Join}
% \vspace{-0.5em}
% \vspace{-1em}

% Reference: 1. (1993 Synchronous Traversal) Efficient processing of spatial joins using R-trees; 2. Simba

Let two datasets $R \subset \mathbb{R}^d$ and $S \subset \mathbb{R}^d$ consist of spatial objects in a $d$-dimensional space. A spatial join between $R$ and $S$ returns a set of all object pairs that meet a specific join predicate, such as \textit{intersects} or \textit{contains}. Equation~\ref{eq:spatial_join} defines spatial join based on the \textit{intersects} predicate. In practice, spatial joins are typically carried out in two or three-dimensional spaces, in which each object can have an arbitrary geometry, such as a point, circle, rectangle, etc.

\vspace{-1em}
\begin{equation}
\label{eq:spatial_join}
R \Join _{\cap } S = \{(r,s) | r \in R, s \in S, r \cap s \neq \varnothing\}
\end{equation}

% * Range query
% * kNN query
% * Spatial join
% * Distance join
% * kNN join

A spatial join involves two steps: \textit{filtering} and \textit{refinement}, and this paper focuses on filtering.

\textbf{Filtering}. The filtering step aims to quickly identify potential candidate pairs of geometries that may satisfy the spatial predicate. 
To achieve this, the filtering step typically relies on either indexes like R-trees~\cite{guttman1984r} and quadtrees~\cite{samet1985storing} or data partitioning to prune non-relevant geometries efficiently and (b) approximate geometries with simpler structures such as \textit{minimum bounding rectangles (MBRs)}. The filtering step returns the set of all object pairs from the two input sets whose MBRs intersect. This step may yield some false positives that satisfy the join predicate on approximated geometries but actually do not upon closer examination.

\textbf{Refinement.} The refinement step takes the results from the filtering step and verifies the actual spatial relationship between the candidate pairs~\cite{brinkhoff1994multi, zimbrao1998raster, georgiadis2023raster}. To eliminate the false positives from the filtering step, this phase determines whether the pairs satisfy the given join predicate (e.g., intersects, contains, etc.) using the actual geometries instead of the MBRs.
% (e.g., intersects, contains, within, equals, touches, crosses, etc.)

% \vspace{-1em}
\subsection{R-tree and Synchronous Traversal}
% \vspace{-0.5em}

Extensive research has been conducted on spatial join algorithms. One category of approaches indexes or partitions one dataset and iterates over all objects of the other dataset as queries. Some representative indexes in this category include the R-tree~\cite{guttman1984r}, linearized KD-Trie~\cite{orenstein1984class}, and simple grid~\cite{vsidlauskas2009trees}.
Another category of approaches simultaneously evaluates both datasets. These approaches include plane-sweep~\cite{de1997computational}, partition-based spatial-merge join (PBSM)~\cite{patel1996partition}, and R-tree synchronous traversal~\cite{brinkhoff1993efficient}.
In this section, we delve into R-tree synchronous traversal due to its great performance~\cite{sowell2013experimental}.

\begin{figure}[t]
	% full width, can be adjusted
  \centering
  \includegraphics[width=0.5\linewidth]{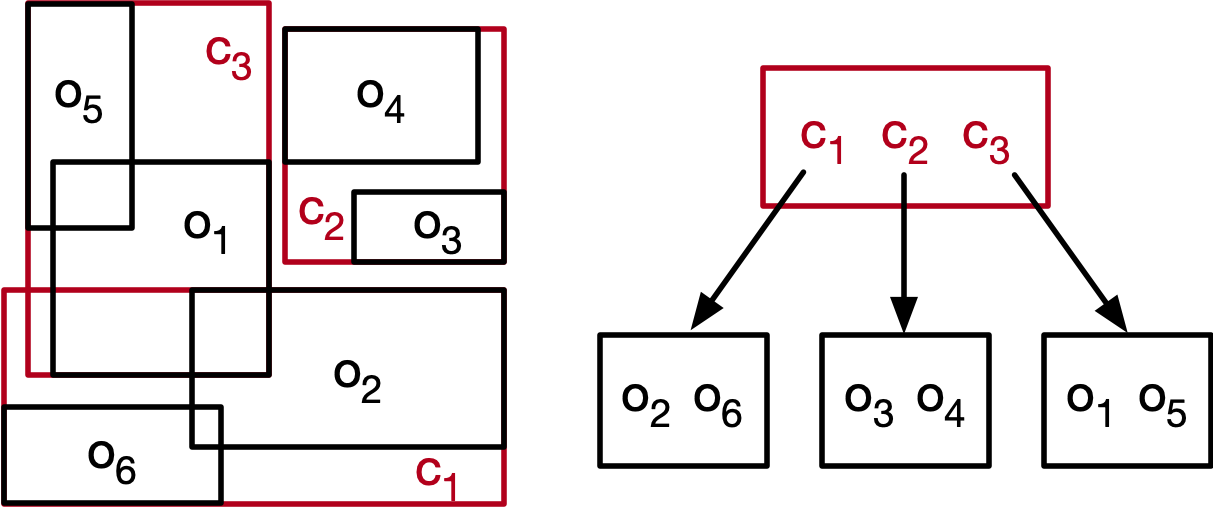}
  % \vspace{-1em}
  \caption{A two-level R-tree index containing six objects.}% given a single client.}
  \vspace{-1em}
  \label{fig:rtree}
\end{figure}

As visualized in Figure~\ref{fig:rtree}, an R-tree is a balanced tree structure. The tree has two types of nodes: directory nodes and leaf nodes. A directory node stores the MBRs and pointers to its child nodes, while a leaf node contains the MBRs and IDs of the spatial objects within the database.

An R-tree satisfies the following properties, considering that $m$ and $M$ represent the minimum and the maximum entries per node, respectively ($2 \leq m \leq \lceil M/2 \rceil$). First, all nodes, excluding the root, contain between $m$ and $M$ entries. Second, the root node may hold only one entry if the root is a leaf node and there is only one object in the dataset, while the upper bound $M$ still applies. Third, all the leaf nodes have the same distance to the root node.

An R-tree can be constructed dynamically or statically. The original R-tree paper~\cite{guttman1984r} manages tree construction by progressively inserting new objects into the tree. The R$^{*}$-tree~\cite{beckmann1990r} enhances tree quality through more intricate insertion strategies. The advantage of dynamic insertion is its ability to intermix with queries and deletions. 
An alternative approach is to bulk-load the entire dataset to construct the R-tree, and the most widely-used approaches are Sort-Tile-Recursive (STR)~\cite{leutenegger1997str} and Hilbert R-tree~\cite{kamel1993hilbert}. Bulk-loading produces superior R-tree topologies compared to dynamically constructed R-trees, improving query performance.

R-tree facilitates efficient spatial query processing by exploiting the hierarchical structure and MBRs to prune the search space. 
% When a spatial query, such as a range or nearest neighbor query, is performed, the tree is traversed from the root node, and only the nodes with MBRs that intersect the query region are examined.
For instance, a window query on R-trees using an intersection predicate seeks to return all objects that intersect with the query geometry.
R-tree handles such a query by starting the search from the root node and traversing the tree down to the leaf nodes. Given a directory node on the search path, all the MBRs of its children are evaluated to determine whether they intersect the query window. Only when an intersection is found will the traversal continue to explore the corresponding subtree. The objects on the searched leaf nodes are compared with the query window, and the qualified ones are returned as results. 

Although a spatial join can be implemented by querying a dataset indexed by an R-tree with many window queries representing the objects from the other dataset, synchronous traversal using R-trees constructed on both datasets can yield better performance~\cite{sowell2013experimental, brinkhoff1993efficient}.
% The following section discusses synchronous traversal in detail.

% {\color{red} introduce window query somewhere?}

\begin{algorithm}[t]
\SetAlgoLined
\SetAlgoNoEnd % Disable 'end' keywords
\caption{Synchronous Traversal}
\SetKwInOut{Input}{input}
\SetKwInOut{Output}{output}
% \KwResult{return all object pairs $\{(r,s) | r \in R, s \in S, r \cap s \neq \varnothing\}$}

\Input{\hspace{0em} RTree\textsubscript{R}, RTree\textsubscript{S} }
\Output{\hspace{0em}  Join results $\{(r,s) | r \in R, s \in S, r \cap s \neq \varnothing\}$}
% \Output{\hspace{0em}  Spatial join result $\{(r,s) | r \in R, s \in S, r \cap s \neq \varnothing\}$}

% ResultSet = RecursiveTraverse(R.root, S.root))
% ResultSet = RecursiveTraverse(RTree\textsubscript{R}.root, RTree\textsubscript{S}.root))

\KwRet{RecursiveTraverse(RTree\textsubscript{R}.root, RTree\textsubscript{S}.root))}
% \KwRet{ResultSet}
\label{algo:sync_traverse}
\end{algorithm}
% \vspace{-1.5em}

\begin{algorithm}[t]
\SetAlgoLined
\SetAlgoNoEnd % Disable 'end' keywords
\caption{Recursive Traverse}
\SetKwInOut{Input}{input}

\Input{\hspace{0em} Node\textsubscript{R}, Node\textsubscript{S}, ResultSet}

\If{Node\textsubscript{R}.isLeaf {\bf and} Node\textsubscript{S}.isLeaf} {
    \For {Obj\textsubscript{R} $\in$ Node\textsubscript{R}, Obj\textsubscript{S} $\in$ Node\textsubscript{S}} {
        \If{Obj\textsubscript{R}.MBR $\cap$ Obj\textsubscript{S}.MBR} {
            ResultSet.add((Obj\textsubscript{R}, Obj\textsubscript{S}))
        }
    }
} 
\ElseIf { Node\textsubscript{R}.isDirectory {\bf and} Node\textsubscript{S}.isDirectory} {
    \For {Child\textsubscript{R} $\in$ Node\textsubscript{R}, Child\textsubscript{S} $\in$ Node\textsubscript{S}} {
        \If{Child\textsubscript{R}.MBR $\cap$ Child\textsubscript{S}.MBR} {
            TraverseRecursive((Child\textsubscript{R}, Child\textsubscript{S}))
        }
    }
}
% \tcp{One node is directory while the other is leaf}
\Else  {
    \For {Child\textsubscript{Dir} $\in$ Node\textsubscript{Dir}} {
        \If{Node\textsubscript{Leaf}.MBR $\cap$ Child\textsubscript{Dir}.MBR} {
            TraverseRecursive((Node\textsubscript{Leaf}, Child\textsubscript{Dir}))
        }
    }
}
% \ElseIf{Node\textsubscript{R}.isLeaf {\bf and} Node\textsubscript{S}.isDirectory} {
%     \For {Child\textsubscript{S} $\in$ Node\textsubscript{S}} {
%         \If{Node\textsubscript{R}.MBR $\cap$ Child\textsubscript{S}.MBR} {
%             TraverseRecursive((Node\textsubscript{R}, Child\textsubscript{S}))
%         }
%     }
% }
% \ElseIf{Node\textsubscript{R}.isDirectory {\bf and} Node\textsubscript{S}.isLeaf} {
%     \For {Child\textsubscript{R} $\in$ Node\textsubscript{R}} {
%         \If{Child\textsubscript{R}.MBR $\cap$ Node\textsubscript{S}.MBR} {
%             TraverseRecursive((Child\textsubscript{R}, Node\textsubscript{S}))
%         }
%     }
% }
\KwRet{}
\label{algo:traverse_recursive}
% \vspace{-1em}
\end{algorithm}
% \vspace{-2em}

% \subsubsection{Synchronous Traversal on R-trees}
% \hfill \\

Synchronous traversal is an efficient spatial join algorithm~\cite{brinkhoff1993efficient}. It refers to the simultaneous traversal of two R-trees, one for each dataset being joined. The goal is to identify object pairs that potentially satisfy the join predicate by exploiting (a) the hierarchical structure of both R-trees and (b) MBRs associated with each R-tree node, such that the search space is pruned effectively.

Algorithms~\ref{algo:sync_traverse} and \ref{algo:traverse_recursive} summarize synchronous traversal on two polygon datasets using the intersect predicate. Starting at the root nodes of both R-trees, as shown in Algorithm~\ref{algo:sync_traverse}, it proceeds in a depth-first search manner.
Algorithm~\ref{algo:traverse_recursive} details the procedure at each traversal step. If both inputs are leaf nodes, the object pairs satisfying the join predicate are added to the result set. If both inputs are directory nodes, the algorithm continues to explore the children pairs that satisfy the join predicate through recursive traversal of the qualified child nodes. If only one of the nodes is a leaf, the recursive traversal proceeds between the leaf node and the qualified children of the directory nodes.

\subsection{Partition-Based Spatial-Merge (PBSM)}
\label{sec:pbsm}

% {\color{red} TODO}

Algorithm~\ref{algo:pbsm} shows the procedure of PBSM. It first partitions input datasets using a uniform grid. Spatial objects are then assigned to the grid tiles they intersect. For each tile, objects from one dataset are joined with those from the other. To avoid duplicate join results, a pair of intersecting rectangles is reported only if a reference point (e.g., the top-left corner) of the intersection region lies within the tile~\cite{dittrich2000data}. 
This partitioning and tile-wise join approach is well-suited for parallel architectures, as demonstrated on multi-core CPUs~\cite{tsitsigkos2019parallel}.

The plane sweep algorithm~\cite{de1997computational} is often applied for tile-wise joins in PBSM instead of the nested loop join used for R-tree node pairs, as plane sweep avoids all-to-all object comparisons. As shown in Algorithm~\ref{algo:plane_sweep}, it first sorts the objects of both input datasets along one axis. A sweep line then moves across the other axis, maintaining an active set of objects that intersect with the line for each dataset. At each step, the algorithm compares the left-most object from the two datasets and inserts it into its respective active set. The algorithm then removes any objects from the opposite active set that no longer intersect with the current sweep line. Afterward, it performs intersection checks between the current object and the remaining objects in the opposite active set. This process repeats until all objects from both datasets have been processed.
By efficiently managing these active sets and only comparing objects that are currently intersecting with the sweep line, the plane sweep algorithm can significantly reduce the number of comparisons compared to the nested loop join.

\begin{algorithm}[t]
\SetAlgoLined
\SetAlgoNoEnd % Disable 'end' keywords
\caption{Partition-Based Spatial-Merge (PBSM)}
\SetKwInOut{Input}{Input}
\SetKwInOut{Output}{Output}

\Input{Dataset $R$, Dataset $S$, Grid $G$ }
% \Output{Join results: $\{(r, s) \,|\, r \in R, s \in S, r \cap s \neq \varnothing\}$}

\tcc{\textbf{Phase 1: Data Partition}}
\ForEach{tile $T_i \in G$} {
    $(R_i, S_i) \leftarrow (R \cap T_i, S \cap T_i)$ 
    % \tcc*[r]{Assign objects in $R$ and $S$ to tile $T_i$}
}

\tcc{\textbf{Phase 2: Tile-wise Join}}
\ForEach{tile $T_i \in G$} {
    ResultSet $\leftarrow$ ResultSet $\cup$ Join($R_i$, $S_i$)\;
}

\KwRet{ResultSet}
\label{algo:pbsm}
\end{algorithm}

\begin{algorithm}[t]
\SetAlgoLined
\SetAlgoNoEnd % Disable 'end' keywords
\caption{Plane Sweep}
\SetKwInOut{Input}{input}

\Input{\hspace{0em} Dataset $R$, Dataset $S$, ResultSet}

\tcc{\textbf{Step 1: Sorting}}
R\textsubscript{sorted}, S\textsubscript{sorted} $\leftarrow$ SortLeft(R), SortLeft(S)\

\tcc{\textbf{Step 2: Sweeping through both datasets}}
ActiveSet\textsubscript{R}, ActiveSet\textsubscript{S}, ResultSet $\leftarrow$ $\emptyset$, $\emptyset$, $\emptyset$\

\While{R\textsubscript{sorted}.notEmpty {\bf or} S\textsubscript{sorted}.notEmpty} {
    
    \If{R\textsubscript{sorted}.first < S\textsubscript{sorted}.first} {
        ActiveSet\textsubscript{R}.insert(R\textsubscript{sorted}.first)
        
        ActiveSet\textsubscript{S}.remove\_inactive(R\textsubscript{sorted}.first)
        
        ResultSet.add(ActiveSet\textsubscript{S}.search(R\textsubscript{sorted}.first))
        
        R\textsubscript{sorted}.pop()
    }
    \Else {
        ActiveSet\textsubscript{S}.insert(S\textsubscript{sorted}.first)
        
        ActiveSet\textsubscript{R}.remove\_inactive(S\textsubscript{sorted}.first)
        
        ResultSet.add(ActiveSet\textsubscript{R}.search(S\textsubscript{sorted}.first))
        
        S\textsubscript{sorted}.pop()
    }
}

\KwRet{ResultSet}
\label{algo:plane_sweep}
\end{algorithm}

% \vspace{-1em}
\subsection{Existing Performance Optimizations}
% \subsection{Existing Performance Optimizations for Spatial Join}
% \vspace{-0.5em}

Over the past four decades, researchers have extensively studied spatial join algorithms, thus, further improvements by algorithm innovations have become increasingly challenging. 
Therefore, to enhance spatial join performance, system implementations have become increasingly important~\cite{vsidlauskas2014spatial}.
This can be achieved by either increasing system resources allocated to the join or boosting efficiency per compute unit through hardware acceleration.

\textbf{Approach 1: scale-up and scale-out.}
Scaling up spatial join on a single server involves harnessing the power of multi-core CPUs and SIMD instructions~\cite{ogden2016gis, rayhan2023simd}.
For further system performance improvements, one can use big data frameworks for spatial data to scale out spatial join operations~\cite{pandey2018good, eldawy2015era}, such as SpatialSpark~\cite{you2015large}, LocationSpark~\cite{tang2016locationspark},  SpatialHadoop~\cite{eldawy2015spatial}, HadoopGIS~\cite{aji2013hadoop}, and Simba~\cite{xie2016simba}.
However, neither the scale-up nor the scale-out solution improves the spatial join efficiency, as the increased performance comes at the expense of additional computing resources.

\textbf{Approach 2: hardware acceleration.}
Researchers have been exploring the use of hardware accelerators such as GPUs to speed up spatial joins. 
GPUs are well-suited for some types of spatial queries, as their graphics primitives operating on polygons align closely with spatial objects~\cite{doraiswamy2020gpu, doraiswamy2022spade}. Specifically, the GPU can map polygons into pixels through rasterization~\cite{zacharatou2017gpu, sun2003hardware} and perform spatial operations, such as intersection checks, at the pixel level. This approach is particularly effective for the refinement step in the spatial join pipeline, as GPUs have significant advantages over CPUs when dealing with highly complex polygons.

However, GPUs have not been shown to be suitable for the filtering step of spatial joins, because the complex control flow in synchronous traversal does not align well with the many-core GPU architecture. 
Even for simpler cases of batched window or range queries on R-trees~\cite{prasad2015gpu, you2013parallel, kim2013parallel, luo2012parallel}, GPUs face significant challenges regardless of whether adopting BFS or DFS. In DFS, each GPU thread is responsible for a single query window, leading to significant workload imbalances between threads, thus the performance is limited by the slowest thread. In BFS, memory management becomes challenging since the number of result pairs produced by a thread block is unknown beforehand. Neither memory overflow nor reserving a huge amount of memory per thread block is ideal. The former leads to execution failures while the latter limits the number of concurrently running thread blocks.

\subsection{Field Programmable Gate Arrays (FPGAs)}

In this work, we prototype our SwiftSpatial accelerator design on FPGAs, leveraging their reconfigurability and widespread availability in data centers.

\textbf{Architecture.} FPGAs lie between general-purpose processors (e.g., CPUs) and application-speciﬁc integrated circuits (ASICs). They behave like ASICs but can be reconfigured virtually an infinite number of times, offering both high performance and design flexibility.
Developers can design custom micro-architectures tailored to specific applications, compile the design to a bitstream file representing the circuit configuration, and load the bitstream onto FPGAs to begin accelerating the applications. 
This process is enabled by a hardware compiler, which maps a logical design to the physical hardware units on FPGAs. These units mainly include (a) Block-RAM (BRAM) and the Xilinx-specific Ultra-RAM (URAM) as small yet fast on-chip memory, (b) Flip-Flops (FF) as registers, (c) Digital Signal Processors (DSP) as computing units, and (d) lookup-tables (LUT) as memory or computing units.

\textbf{Programming.} Developing FPGA accelerators typically requires significantly more effort compared to software design. Traditionally, FPGAs are programmed using hardware description languages (HDL), such as Verilog and VHDL, where developers define the circuit behavior at the granularity of a single clock cycle. Recently, High-Level Synthesis (HLS) has gained popularity, as it allows programmers to design circuits at a higher level using C/C++ or OpenCL~\cite{vivado_hls, intel_opencl_fpga}. However, developing accelerators with HLS demands additional effort to learn the specific hardware-friendly coding style and to fine-tune performance by exploring a wide range of \textit{pragmas}.

\section{SwiftSpatial: Accelerator Design}
\label{sec:accelerator}
% \vspace{-0.5em}

We propose \textit{SwiftSpatial}, an FPGA-based accelerator for high-performance and energy-efficient spatial join filtering. We introduce the hardware design in this section, leaving its system integration in the following section.

% In this section, we first outline the design principles for an efficient spatial join accelerator~(\S\ref{sec:design_principles}) and overview SwiftSpatial~(\S\ref{sec:accelerator_overview}). 
% Subsequently, we delve into the key components of SwiftSpatial, including the join units~(\S\ref{sec:join_units}), the on-chip scheduler~(\S\ref{sec:scheduler}), and the memory management units~(\S\ref{sec:memory_management}). 
% Next, we describe SwiftSpatial's FPGA implementation~(\S\ref{sec:FPGA_implementation}). 
% Finally, we discuss the approaches to incorporate SwiftSpatial into spatial data management systems~(\S\ref{sec:integration}).

% \vspace{-1em}
\subsection{Design Principles}
\label{sec:design_principles}
% \vspace{-0.5em}
 
As introduced in \S\ref{sec:background}, spatial join algorithms commonly decompose the join process into many sub-join tasks on tiles of objects (e.g., nodes in R-trees or tiles in PBSM). 
\textbf{The key idea of SwiftSpatial is to offer fast hardware primitive for joining small tiles of objects.}
Consequently, an efficient accelerator for spatial joins should meet three fundamental criteria: 
\begin{itemize}
% \vspace{-0.5em}
    \item Swift execution of each sub-join on a tile pair.
% \vspace{-0.5em}
    \item Parallelization of sub-join tasks on multiple join units.
% \vspace{-0.5em}
    \item Minimal task parallelization overhead even for fine-grained control flows like R-tree synchronous traversal.
\end{itemize}
% \vspace{-1em}

% \vspace{-1em}
\subsection{Accelerator Overview}
\label{sec:accelerator_overview}
% \vspace{-0.5em}

\begin{figure}[t]
	% full width, can be adjusted
  \centering
  \includegraphics[width=0.8\linewidth]{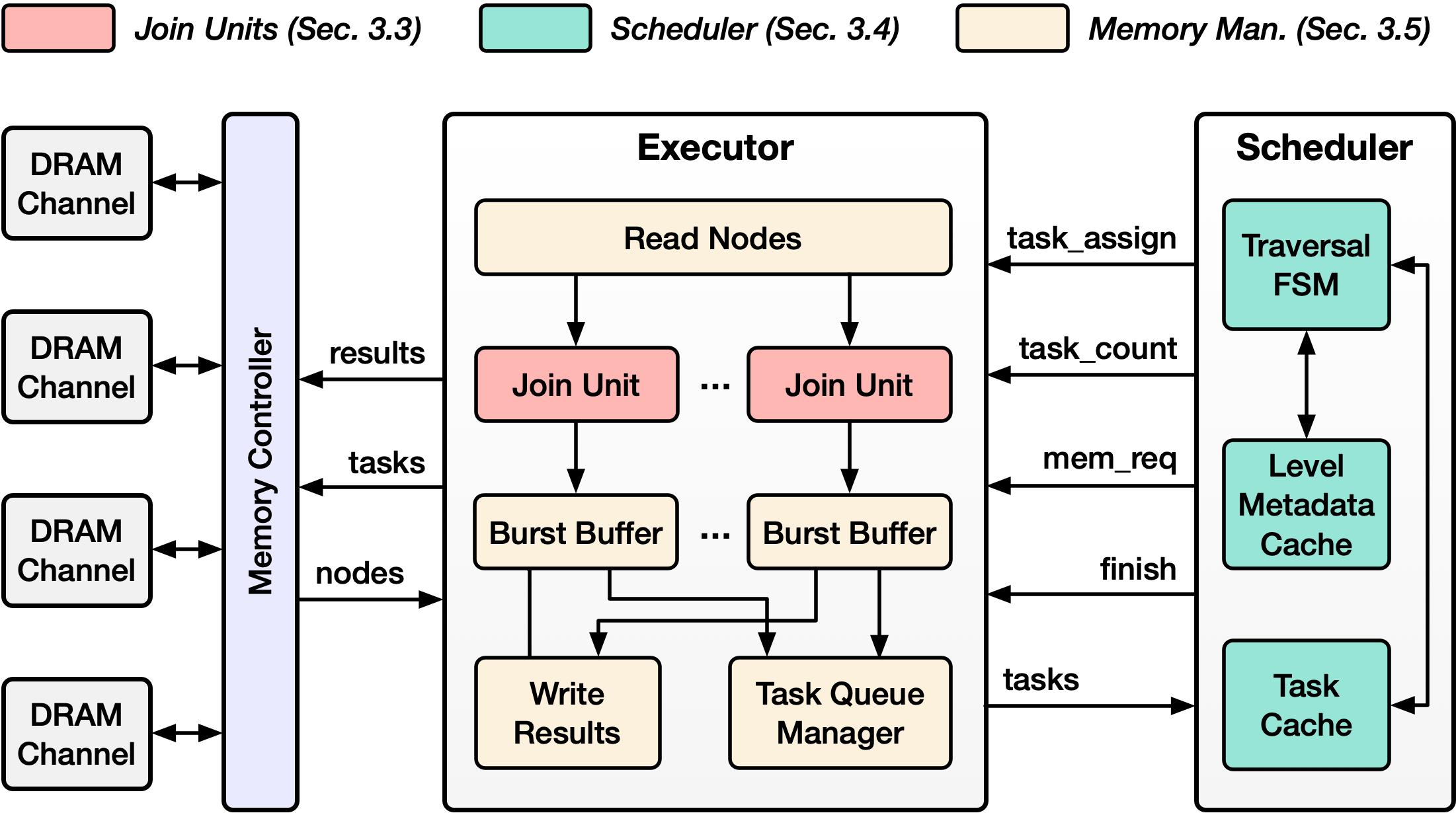}
  % \vspace{-1em}
  \caption{SwiftSpatial accelerator overview.}
  \vspace{-1em}
  % \caption{The design of the SwiftSpatial accelerator primarily consists of a task scheduler, a set of identical join units, and various memory management units.} 
  \label{fig:accelerator}
\end{figure}

Reflecting the above criteria, we design \textit{SwiftSpatial}, an in-memory spatial join accelerator for the filtering phase.

SwiftSpatial instantiates multiple specialized join units for the efficient joining of object tiles. 
The join units implement nested loop join rather than plane sweep for the following reasons. 
While plane sweep --- designed to minimize the number of predicate evaluations through sorting --- is a frequent choice in software solutions, this preference stems from the slow predicate evaluations on CPUs. 
In contrast, with our proposed architecture employing hybrid parallelism, predicate evaluations between objects become notably faster, even if each evaluation involves multiple object boundary comparisons. 
Furthermore, the efficiency of plane sweep depends on data distributions and tile sizes. Specifically, its performance often lags behind the nested loop join on datasets with overlapped objects or when utilizing small tiles, as we will illustrate in \S\ref{sec:eval}.
Finally, plane sweep is a sequential algorithm that benefits more from a high clock frequency than from a parallel hardware architecture. 

Another key accelerator component is the on-chip scheduler that manages the join control flow and the coordination of the join units. 
Executing the control flow on-chip results in low performance overhead compared to the alternatives like relying on an external CPU for control signal transmission. In this paper, we not only implement PBSM with a simple control flow but also incorporate R-tree synchronous traversal into the scheduler, showcasing the efficient execution of a complex control flow in SwiftSpatial. The integration of R-tree synchronous traversal also allows SwiftSpatial to leverage the existing R-tree indexes maintained by spatial data management systems, such as PostGIS~\cite{postgis}, Oracle Spatial~\cite{kothuri2002quadtree}, Apache Sedona~\cite{sedona}, and SpatialSpark~\cite{spatialspark}, avoiding additional index construction costs during spatial joins.

% (a) it is widely known for its impressive performance not only on static datasets~\cite{brinkhoff1993efficient} but also on moving-object datasets~\cite{sowell2013experimental}, and (b) R-trees are widely used in spatial databases~\cite{kothuri2002quadtree, xie2022ganos} and spatial big data frameworks~\cite{you2015large, eldawy2015spatial, aji2013hadoop}.
% Our adoption of R-tree as the spatial index is rooted in three considerations. 
% Firstly, it is the de-facto index in various spatial DBMSs, including PostGIS~\cite{postgis}, Oracle Spatial~\cite{kothuri2002quadtree}, Apache Sedona~\cite{sedona}, and SpatialSpark~\cite{spatialspark}. Given that a DBMS already maintains an R-tree, SwiftSpatial can directly leverage it for spatial joins, thereby bypassing any additional index construction.

\textbf{Figure~\ref{fig:accelerator} overviews SwiftSpatial.} Beyond the join units and the on-chip scheduler, SwiftSpatial encompasses a set of memory management units, which optimize data transfer between the accelerator and DRAM. On-chip communication between function units, like the join units and the scheduler, is facilitated by FIFOs, akin to pipes in a software context.

% The outputs of the join units are directed to either the result writing unit (if both input nodes are leaf nodes) or the task queue manager (if at least one input node is a directory node). The node pairs in the task queue are then used as input for further tree traversal.

% synchronous traversal and assigns tasks to all other units (read and write units, join units, task queue manager, etc.).

% The FPGA contains several DRAM channels and a memory controller, as shown on the left side of Figure~\ref{fig:accelerator}. The FPGA DRAM is responsible for holding the R-trees and the results produced by the FPGA. The accelerator features on-chip memory management units that manage and optimize data transfer to the memory controller.
% On-chip communication between function units is facilitated by FIFOs instantiated with BRAMs (a type of FPGA on-chip SRAM), which are similar to pipes in a software context.

% \vspace{-1em}
\subsection{Join Units}
\label{sec:join_units}
% \vspace{-0.5em}

A join unit processes a pair of nodes from the two R-trees being joined and generates all intersected entry pairs.

\begin{figure}[t]
	% full width, can be adjusted
  \centering
  \includegraphics[width=0.8\linewidth]{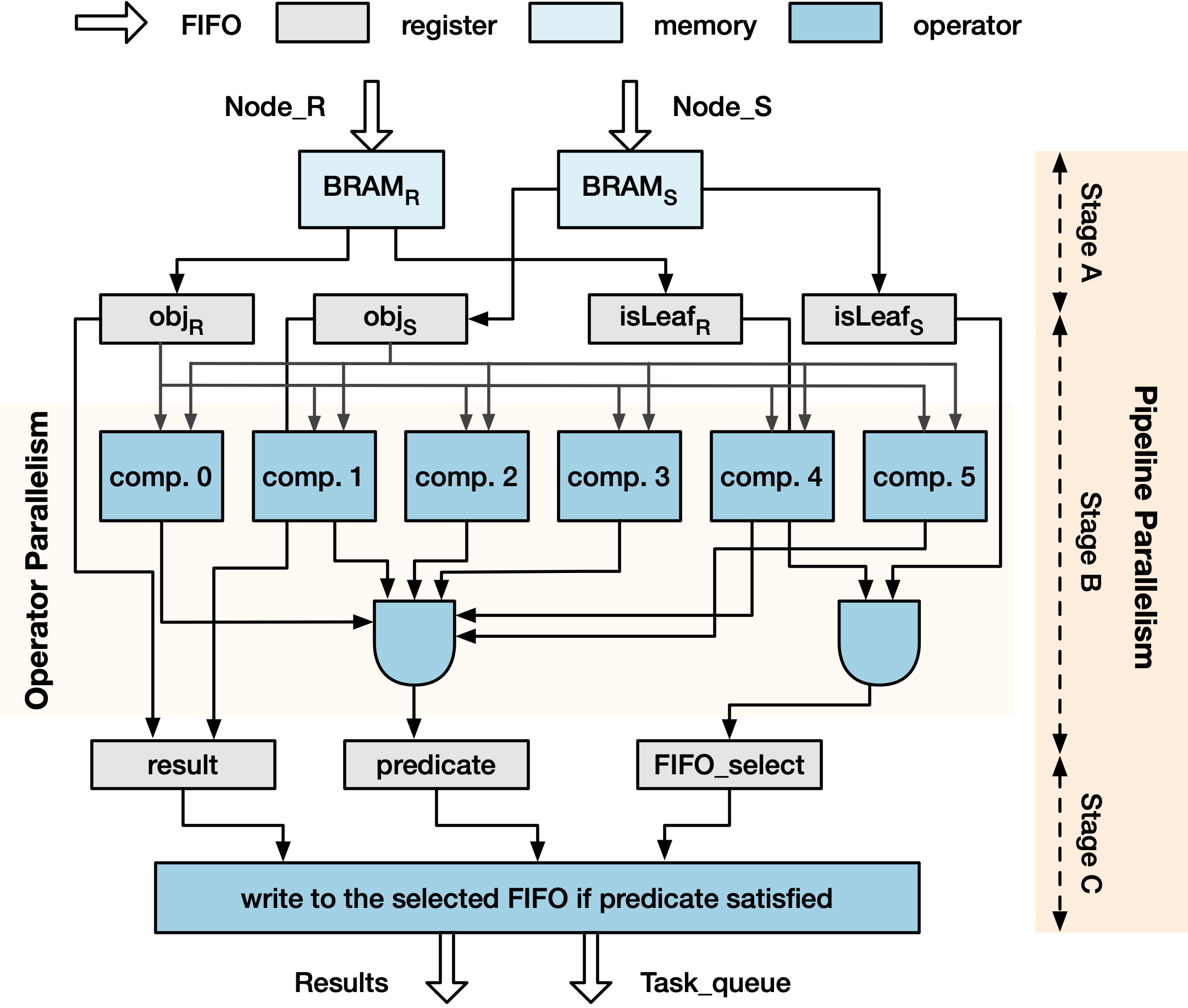}
  % \vspace{-1em}
  \caption{The microarchitecture design of a join unit.}% given a single client.}
  \vspace{-1em}
  \label{fig:join-unit}
\end{figure}

\textbf{Figure~\ref{fig:join-unit} shows the microarchitecture design of a join unit.}
The input node pairs are streamed into the join unit through two FIFOs. 
A join unit consists of two on-chip SRAM slices used to store the pair of input nodes. 
During each iteration of the join process, a pair of objects from the nodes are read from SRAM into registers. Then, multiple comparators are employed to assess whether the MBRs intersect, where all the comparators work in parallel. 
Specifically, there are six comparisons for 3D MBRs or four for 2D MBRs: \textit{r.right}$\ge$\textit{s.left}, \textit{s.right}$\ge$\textit{r.left}, \textit{r.top}$\ge$\textit{s.bottom}, \textit{s.top}$\ge$\textit{r.bottom}, \textit{r.front}$\ge$\textit{s.back}, \textit{s.front}$\ge$\textit{r.back}, where $r$ and $s$ are two entries from the pair of input nodes. 
% Specifically, there are four comparisons for 2-dimensional MBRs: \textit{r.right}$\ge$\textit{s.left}, \textit{s.right}$\ge$\textit{r.left}, \textit{r.top}$\ge$\textit{s.bottom}, \textit{s.top}$\ge$\textit{r.bottom}, where $r$ and $s$ are two entries from the pair of input nodes. 
% For the contain predicate, the four comparisons are: \textit{r.left}$\le$\textit{s.left}, \textit{r.right}$\ge$\textit{s.right}, \textit{r.top}$\ge$\textit{s.top}, and \textit{r.bottom}$\le$\textit{s.bottom}.
% For 2-dimensional MBRs, the third dimension is always padded with zeros at evaluation time, thus the last two checks are always true.
If all the comparison outcomes are true (evaluated by the AND gate), the pair of objects will be written to one of the output FIFOs.
If both nodes are leaf nodes, the pair is written to the final result FIFO; otherwise, it is written to the task queue FIFO.

\textbf{A join unit processes a pair of objects per cycle, thanks to the hybrid parallelism in the architecture.} 
As shown in Figure~\ref{fig:join-unit}, the first level of parallelism is operator parallelism. The combinational logic that evaluates different comparisons simultaneously and aggregates the comparison results can be completed within a clock cycle.
Another form of parallelism is pipeline parallelism.
Figure~\ref{fig:join-unit} shows different levels of registers in the pipeline. In the first stage of the pipeline, the objects are read from SRAM into the object registers. In the second stage, whether the MBRs intersect is evaluated. In the final stage, the qualified results are written into FIFOs. Thus, there are three object pairs in progress within the pipeline, allowing one pair of objects to be introduced into the pipeline every single clock cycle.

\begin{figure}[t]
	% full width, can be adjusted
  \centering
  \includegraphics[width=0.48\linewidth]{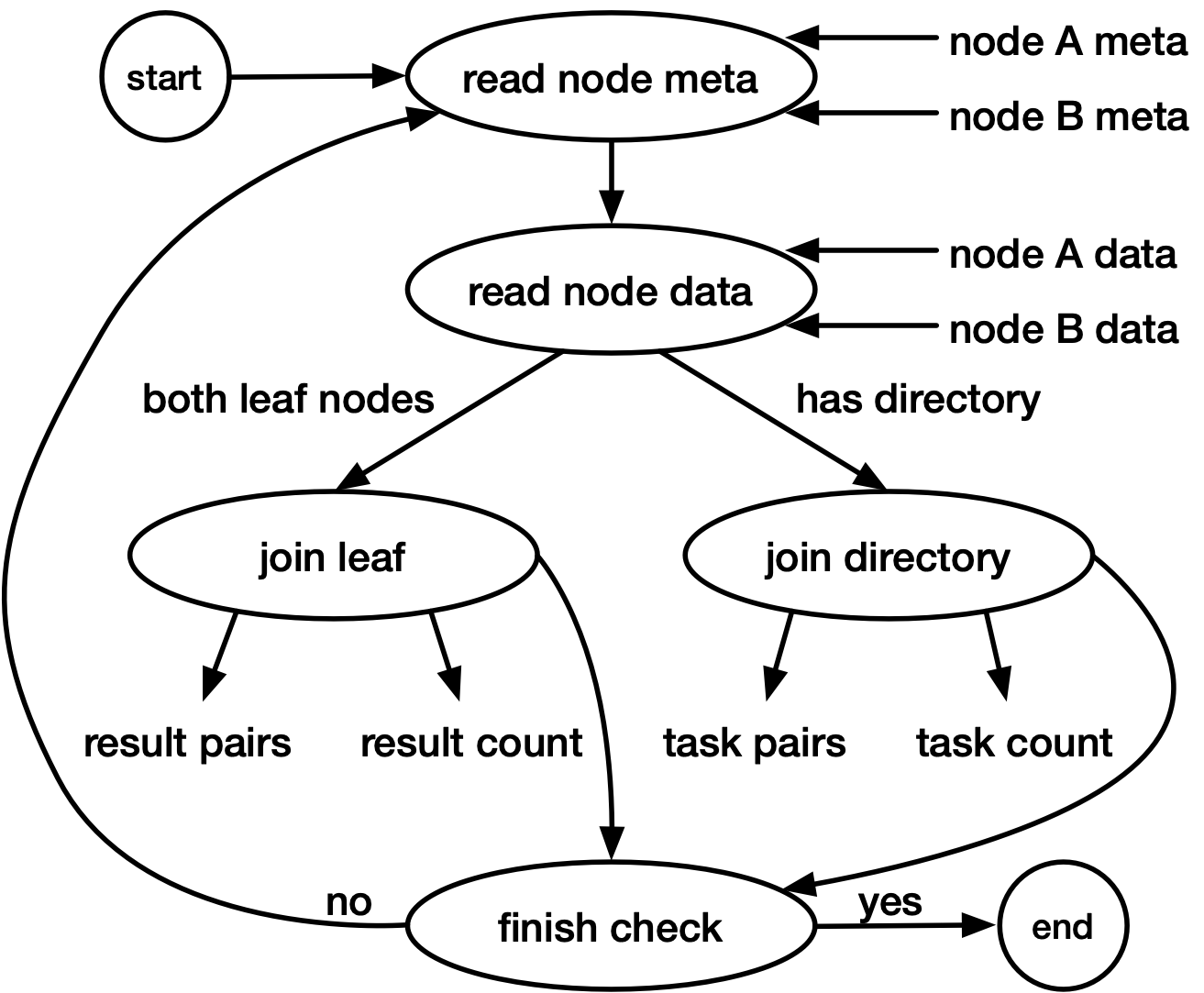}

  % \vspace{-1em}
  \caption{The control flow of a SwiftSpatial join unit.}% given a single client.}
  \vspace{-1em}
  \label{fig:join-unit-fsm}
\end{figure}

Figure~\ref{fig:join-unit-fsm} demonstrates the control flow of a hardware spatial join unit. 
% Figure~\ref{fig:control-fsm}~(a) demonstrates the control flow of a hardware spatial join unit. 
The join unit follows a series of steps until it receives the finish signal. Initially, it receives node metadata from the read unit, which indicates the number of node entries and whether it is a leaf or directory node. Based on this information, the join unit reads the corresponding node data. Once the reading process is complete, the join procedure starts. During the procedure, the join unit examines every pair of entries between the node pair and checks if they intersect, outputting the qualifying pairs. If both input nodes are leaves, the outputs are recorded as results; otherwise, the generated node pairs are treated as tasks for future join operations. After joining the entries, the join unit checks if there is a finish signal sent from the scheduler, and it proceeds with the next pair of nodes if it has not received one.

% \begin{figure}[t]

%   \begin{subfigure}[b]{0.5\linewidth}
%     \includegraphics[width=\linewidth]{fig/join-unit-fsm.png}
%   \end{subfigure}
  
%   \begin{subfigure}[b]{0.5\linewidth}
%     \includegraphics[width=\linewidth]{fig/scheduler-fsm.png}
%   \end{subfigure}
  
%   \begin{subfigure}[b]{0.5\linewidth}
%     \includegraphics[width=\linewidth]{fig/task-queue-manager-fsm.png}
%   \end{subfigure}
%   \caption{The control flow of the join units, on-chip scheduler, and task queue manager.}
  
%   \label{fig:control-fsm}
% \end{figure}
  
% \begin{figure*}[t]
% 	% full width, can be adjusted
%   \centering
%   \includegraphics[width=1.0\linewidth]{fig/three_control_FSMs.png}
%   % \vspace{-1.5em}
%   \caption{\red{TODO}The control flow of the join units, on-chip scheduler, and task queue manager.}% given a single client.}
%   % \vspace{-1em}
%   \label{fig:control-fsm}
% \end{figure*}

% \vspace{-1em}
\subsection{On-chip Scheduler} 
\label{sec:scheduler}
% \vspace{-0.5em}

Although one approach to manage spatial join on SwiftSpatial is to combine the hardware join primitives with a software control flow for task scheduling, this method can be time-consuming, because it necessitates many accelerator kernel invocations and frequent data transfers between the host server and the accelerator.

To avoid the software scheduling overhead, SwiftSpatial incorporates an on-chip hardware scheduler to manage the control flow of spatial join and coordinate all function units, including the read and write units, join units, and the task queue manager.

\subsubsection{Synchronous Traversal Scheduler} 

\textbf{We modify the synchronous traversal algorithm to adopt BFS, maximizing task-level parallelism among the join units.} This is essential because the original synchronous traversal, as illustrated in Algorithm~\ref{algo:traverse_recursive}, employs a depth-first search (DFS), which allows joining only one node pair at a time and fails to fully leverage the parallelism capabilities of SwiftSpatial. 
By contrast, the BFS synchronous traversal~\cite{huang1997spatial} performs the join level by level, such that the tasks in each level can be easily parallelized. At each level, the scheduler considers all pairs of nodes to join and assigns the tasks to the join units. For instance, at the root level, the sole pair of nodes to join is the pair of roots. The scheduler assigns the join task to the first join unit and waits for the result to be written to memory by the task queue manager. Assuming the roots are not leaf nodes, all the results generated by joining the roots become tasks for the second level, and the scheduler assigns these tasks to the join units. This process continues until all tasks at the leaf level are completed, and the results are written back to memory.

\begin{figure}[t]
	% full width, can be adjusted
  \centering
  \includegraphics[width=0.62\linewidth]{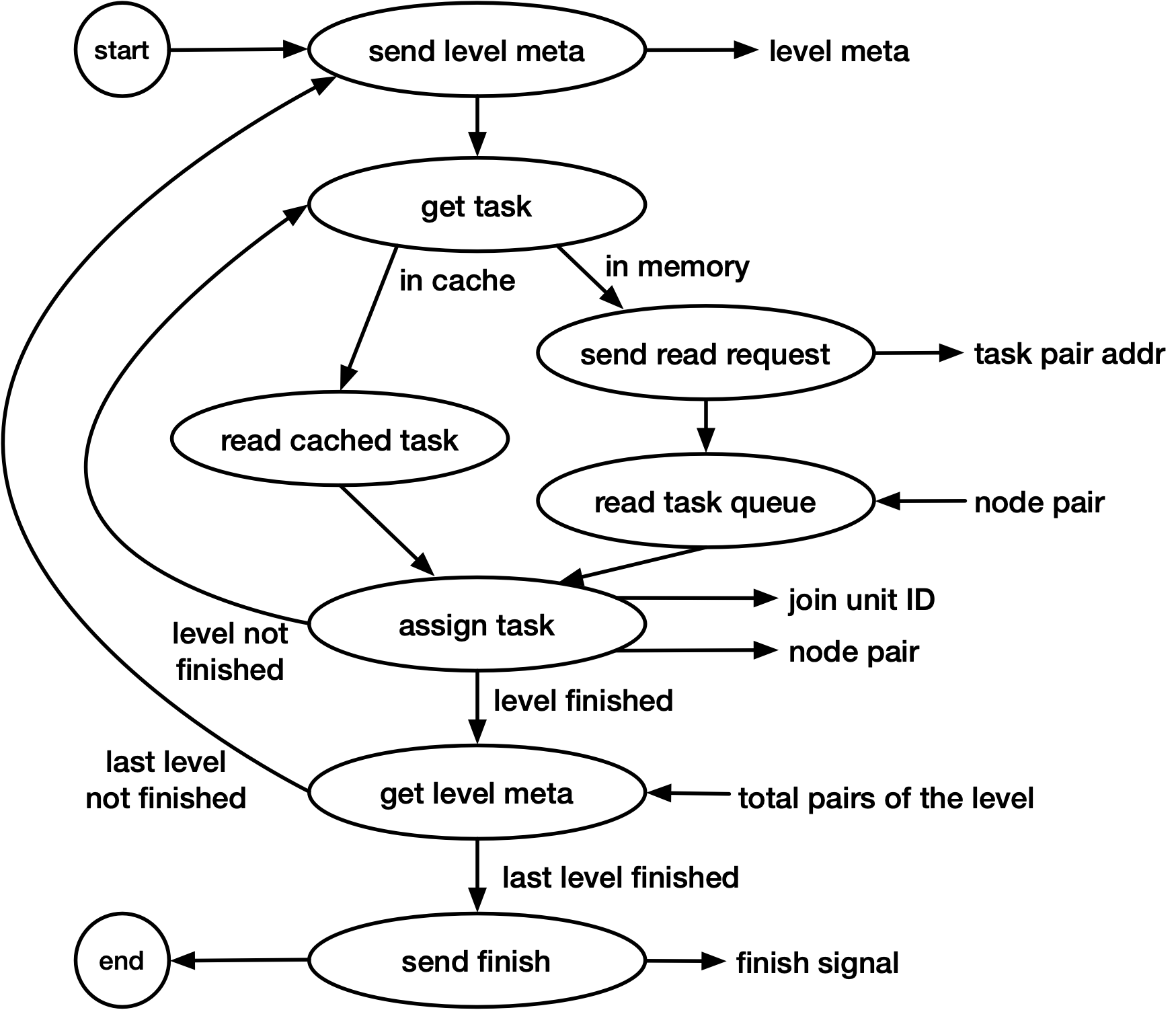}
  % \vspace{-1em}
  \caption{The control flow of the on-chip scheduler.}% given a single client.}
  \vspace{-1em}
  \label{fig:scheduler-fsm}
\end{figure}

% Figure~\ref{fig:control-fsm}~(b) details the workflow of the scheduler. 
Figure~\ref{fig:scheduler-fsm} details the workflow of the scheduler. 
To start the BFS synchronous traversal in a level, the scheduler sends the physical addresses to the task queue manager to inform it where to begin writing the intermediate results (future tasks).
Next, it starts dispatching jobs to the join units.
To accomplish this, it requests the results from the previous level from the task queue manager. 
Because the task queue manager needs to read the requested task (a pair of nodes) from memory, which involves a random DRAM access, we optimize memory bandwidth utilization by implementing burst loading: reading a sequence of node pairs that can be issued next and caching them in the scheduler's SRAM. 
As a result, the scheduler only needs to send a request to the task queue manager when its local cache is empty. 
After obtaining the node pair as a task, the scheduler looks for a join unit to assign the task. Specifically, the scheduler dispatches tasks to the join units in a round-robin manner.
With the node pair and join unit ID, the scheduler sends this information to the read unit, which loads the node pair from memory and passes the data to the respective join unit.
The task assignment continues until all the tasks on the current level are completed. 
The task queue manager then sends the scheduler the number of pairs produced in the current level, allowing the scheduler to determine the starting physical address to write from in the next level.
The aforementioned procedures are repeated until the leaf level is completed. The scheduler finally sends a finish signal to other function units, indicating all tasks have been dispatched.

\subsubsection{PBSM Scheduler} 

The PBSM scheduler operates similarly to the leaf-level join in synchronous traversal. Given multiple tiles to be joined, the scheduler distributes tasks across different join units. SwiftSpatial supports both static and dynamic scheduling policies. In static scheduling, tasks are predefined and assigned to specific join units, while in dynamic scheduling, an incoming task is allocated to the first available idle join unit. Empirically, these scheduling policies show similar performance, as the large number of tiles to be joined typically balances workload distribution across the join units.

We adjust the partitioning strategy on the software side to better leverage the join units. Since each join unit performs a nested loop join instead of a sweep, join performance is sensitive to tile size, as the number of object pairs to compare grows quadratically with the geometric mean of the number of objects in a tile. Thus, we set an upper bound of workload per tile by allowing hierarchical partitioning for tiles containing a large number of objects. For example, with a geometric mean tile size of 32, the number of comparisons to perform between the objects of the two datasets is limited to $32^2=1024$, although one dataset may contain more than 32 objects in the tile.

\subsection{Memory Management}
\label{sec:memory_management}

\textbf{Physical address space management.} 
SwiftSpatial directly manages the physical address space, eliminating the overhead associated with dynamic memory allocation, deallocation, and virtual address space management typically seen in CPU- and GPU-based spatial joins.
For write operations, SwiftSpatial uses a self-incrementing counter to efficiently manage result writing: intermediate or final results from all join units are streamed into the write units, which concatenate the results and write them to memory in bursts.
This approach removes the need for individual join units to allocate memory based on the join cardinality.

\textbf{Optimizing bandwidth utilization via request bursting.}
While reading nodes exhibit a sequential memory access pattern, the write requests in the accelerator can be highly random. This randomness arises from join operations, where, for example, it may take a join unit several cycles to output the second pair of intersected objects after the first is sent. Several join units can interleave write requests, with each write being only 8-byte long (a pair of integer object IDs). 

To solve this problem, we introduce a burst buffer after every join unit. The burst buffer takes the join results as input and sends a burst of results to the task queue manager or the result write unit. The burst buffer will output a burst either when the accumulated results reach a size threshold, e.g., 4 KB, or when it is the end of joining a pair of nodes. The write result unit or the task queue manager pulls the burst results from the burst buffers in a round-robin fashion and writes the data to DRAM sequentially.

\textbf{Task queue manager.}
According to the scheduler's requests, the task queue manager (a) writes the intermediate results, i.e., the node pairs on non-leaf levels that intersect, as future tasks to memory or (b) reads node pairs to be joined from memory.

\begin{figure}[t]
	% full width, can be adjusted
  \centering
  \includegraphics[width=0.52\linewidth]{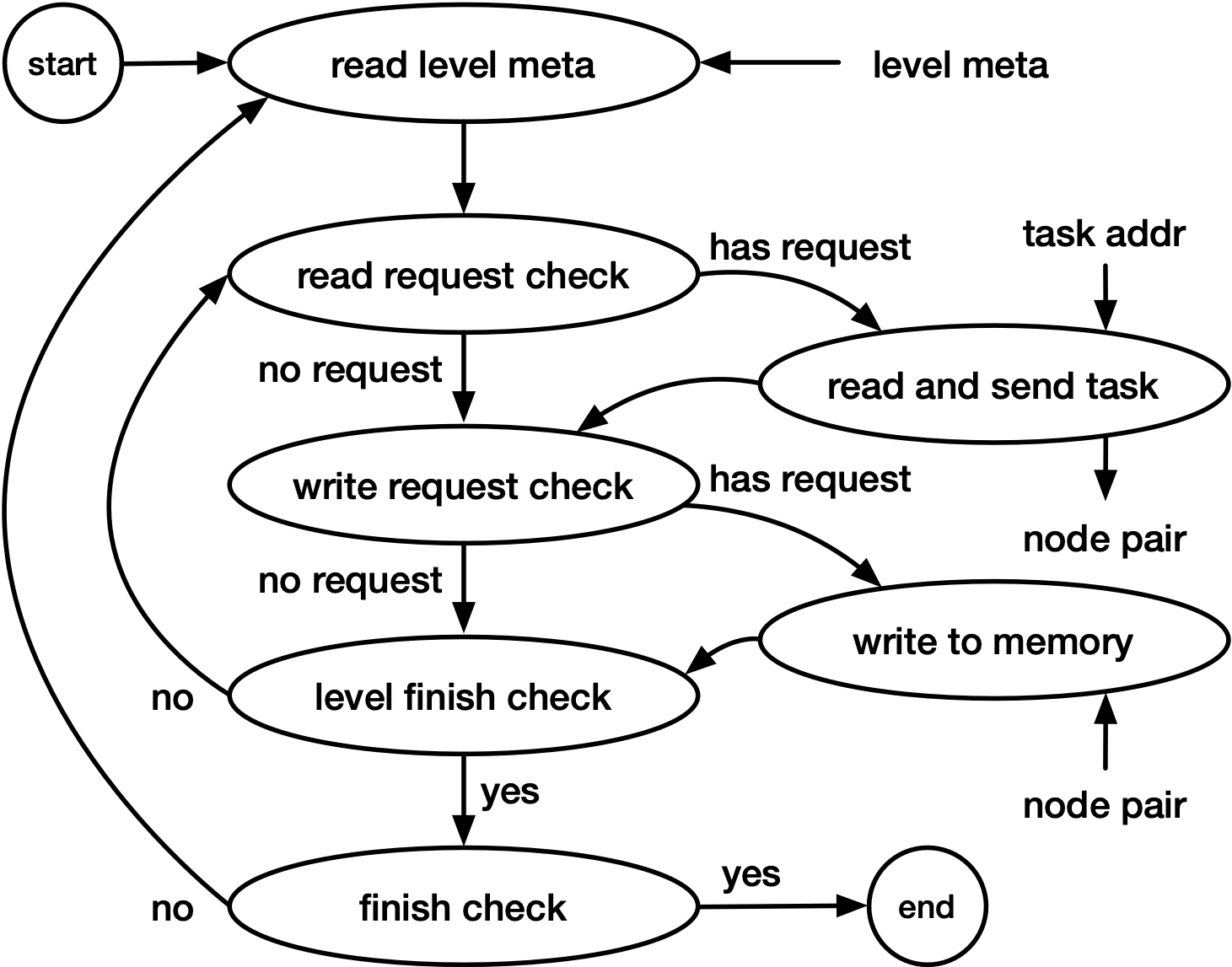}
  % \vspace{-1em}
  \caption{The control flow of the task queue manager.}% given a single client.}
  \vspace{-1em}
  \label{fig:task-queue-manager-fsm}
\end{figure}

% The task queue manager workflow is illustrated in Figure~\ref{fig:control-fsm}~(c).
The task queue manager workflow is illustrated in Figure~\ref{fig:task-queue-manager-fsm}.
It starts by receiving the level metadata from the scheduler, which indicates the number of join tasks in the current level and the starting physical address for storing intermediate results.
Next, it checks for any read requests from the scheduler and returns the corresponding node pairs.
Subsequently, the task queue manager pulls data from the burst buffer and writes these intermediate results to memory in bursts.
The process continues until the manager receives a level finish signal at the leaf level.

\subsection{FPGA Implementation}
\label{sec:FPGA_implementation}

We choose to develop SwiftSpatial on FPGAs for two primary reasons: (a) it facilitates direct performance evaluation of the accelerator in contrast to the simulation-based approach, and (b) the widespread adoption of FPGAs in data centers makes SwiftSpatial readily integrable with spatial data management systems~\cite{firestone2018azure, aqua, li2019cloud, zhang2020fpga, maydash}.
We developed SwiftSpatial on FPGAs using Vitis HLS 2022.1 in C/C++ and set the accelerator's clock frequency as 200 MHz.
% The on-chip memory (SRAM) and FIFOs in SwiftSpatial correspond to BRAMs on the FPGA, supporting a read or write request every clock cycle, while registers are mapped to Flip-Flops (FFs).
% The FPGA's digital signal processors (DSPs) execute floating-point operations, with other logic operations facilitated via lookup tables (LUTs).

% \vspace{-1em} 
\section{System Integration}
\label{sec:integration}
% \vspace{-0.5em}

A spatial data management system could leverage SwiftSpatial in a hybrid CPU-FPGA architecture.
In this architecture, the CPU constructs and maintains spatial indexes (e.g., R-trees) or partitions (e.g., PBSM) and handles non-join queries such as window queries, while the accelerator processes spatial join queries. 
Upon receiving a spatial join query, the up-to-date indexes are transferred to the accelerator, which then processes and returns the join results. As we will show in the experiments, the index transfer time is negligible compared to the computationally intensive join operation.

\subsection{FPGA Placements in the System}

\begin{figure}[t]
	% full width, can be adjusted
  \centering
  \includegraphics[width=0.6\linewidth]{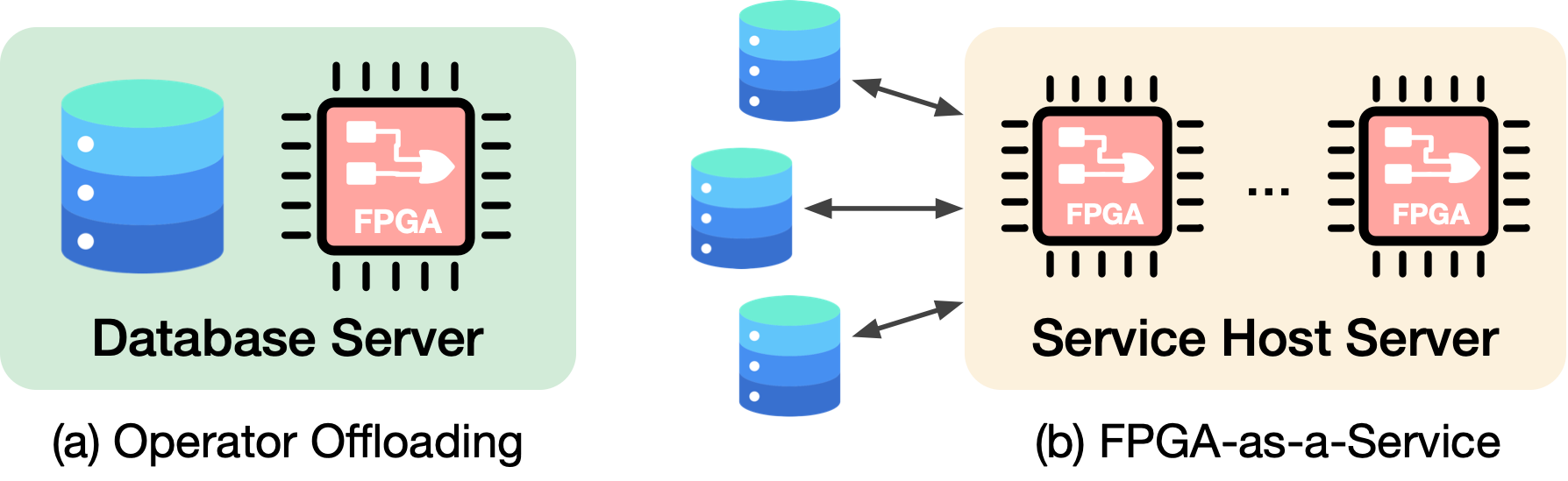}
  \vspace{-.5em}
  \caption{Two ways of integrating SwiftSpatial into a spatial data management system.}% given a single client.}
  \vspace{-1em}
  \label{fig:system-integration}
\end{figure}

There are two ways to integrate SwiftSpatial into a spatial data management system (Figure~\ref{fig:system-integration}).

\textbf{Operator offloading.}
In this approach, the CPU server operates a data management system while the accelerator serves as an offloading engine. An example of this offloading architecture is Alibaba's cloud-native database engine PolarDB, which accelerates LSM-trees (log-structured merge trees) by offloading compactions to FPGAs~\cite{huang2019x, zhang2020fpga}.

\textbf{FPGA-as-a-Service (FaaS).}
SwiftSpatial can function as a cloud-based service accessible to multiple users, avoiding the cost of integrating an accelerator per DBMS.
Users can submit spatial join queries to the service, either by providing raw datasets or serialized R-trees.
The host server of SwiftSpatial processes the user's requests from the network, constructs indexes if needed, forwards the queries to the accelerator, and returns the results to the users. 
An example of accelerator-as-a-service in the cloud is SAP DASH~\cite{maydash}, which presents the FPGA as a hardware data processing service to a cluster of databases, delegating the asynchronous execution of compute-intensive database operations.

\subsection{Request Scheduling and Multi-tenancy}

In the FaaS model, the FPGA host server is responsible for scheduling the requests to the FPGAs. 

Each FPGA can instantiate either one large or multiple smaller SwiftSpatial kernels. Since an FPGA has fixed hardware resources, both methods are bounded by the same limit for resource consumption, thereby offering the same compute capabilities. When a single accelerator kernel is instantiated, user queries are queued and processed sequentially. Alternatively, multiple smaller kernels allow for the concurrent processing of several queries.

The difference between the two approaches lies in memory management and fairness. The multi-tenancy approach enhances fairness, preventing a situation where the accelerator is monopolized by a single query for an extended period, while other users are left waiting. However, this approach requires sharing memory between queries, which can limit the memory available for each query and potentially degrade performance compared to the single kernel instantiation.

% \vspace{-1.5em}
\section{Evaluation}
\label{sec:eval}
% \vspace{-0.5em}

We evaluate SwiftSpatial on real-world and synthesized datasets and compare it to a wide range of CPU- and GPU-based spatial join implementations. We show that SwiftSpatial outperforms the most performant software baseline by up to 41.03$\times$ in terms of latency, while requiring 6.16$\times$ less power. The modular architecture of SwiftSpatial facilitates its deployment not only on data-center-grade FPGAs but also on embedded systems, showcasing its adaptability and potential for broad applicability.

% \vspace{-1em}
\subsection{Experimental Setup}
% \vspace{-0.5em}

\textbf{Datasets.}
We evaluate SwitftSpatial on both real-world and synthesized datasets. For real-world datasets, we use the Open Street Maps (OSM) dataset made available by SpatialHadoop~\cite{spatialhadoop}. Specifically, we process its \textit{buildings} subset for polygon generations (converted to MBRs) and use the \textit{all nodes} subset for point objects. The dataset scales range from one hundred thousand to ten million objects, and we produce two different datasets for the spatial join operation. We also produce synthesized datasets of rectangles in uniform distribution. Specifically, we set the map size as 10K by 10K, within which each unit-square object is randomly and uniformly distributed.

% For Gaussian distribution, we randomly select some centroids where objects surronds in an Gaussian distribution. Specifically, we set the number of centroids as one percent of total objects, and each centroids surrond one hundred objects. 
% Similar to the OSM dataset, we synthesize two datasets per scale in each data distribution.
% For both OSM and Uniform datasets, we evaluate join performance on different data scales ranging from 100 thousand to 10 million objects. 

\textbf{Software.}
For CPU baselines, we evaluate both open-source software and our multi-threaded C++ synchronous traversal and PBSM implementations.

The open-source software baselines include PostGIS~\cite{postgis}, Apache Sedona~\cite{sedona}, and SpatialSpark~\cite{spatialspark}, thus covering both spatial database and spatial big data frameworks. 
PostGIS is a spatial database that extends PostgreSQL with spatial functionalities. 
Apache Sedona, previously known as GeoSpark~\cite{geospark}, is a system for processing large-scale spatial data. Even on a single server, Sedona carries out spatial joins on multiple-core CPUs by performing join on each data partition and subsequently merging the results. 
Similar to Sedona, SpatialSpark is an academic big spatial data framework developed on top of Apache Spark~\cite{zaharia2012resilient}. 
All the CPU baseline software above performs spatial join using R-trees.
For each query, we perform a warmup run followed by three executions, from which we report the performance results. 
Since the datasets fit in main memory, the data already resides in memory during subsequent runs, avoiding data movement overhead from disks.
% except our own implementation using R-tree but handle spatial join by window-queries 
% {\color{red} TODO}

We implement multi-threaded C++ spatial join baselines, including R-tree synchronous traversal and PBSM. 
Our implementations are adapted from the single-threaded artifacts in~\cite{sowell2013experimental}. 
For synchronous traversal, we implement two multi-threaded versions using OpenMP, covering both BFS and DFS.
The first version applies a BFS traversal, identical to the FPGA implementation illustrated in \S\ref{sec:scheduler}. At each level of the BFS traversal, we consider the node pairs for joining as tasks to be assigned to various threads. Each thread is responsible for a subset of tasks, with a single thread subsequently merging the results. 
The second version combines DFS with BFS. The idea is to start the search with multi-threading BFS, and, once the number of node pairs to join surpasses the number of threads significantly, these tasks are dispatched to different threads. These threads manage the tasks using the conventional DFS synchronous traversal, after which a single thread merges the results. Specifically, when the task count is at least ten times greater than the hardware thread number, the algorithm switches to DFS. This is because assigning only one task per thread can lead to significant workload imbalances due to the varying DFS traversal time per task.
For both BFS and BFS-DFS implementations, we evaluate two OpenMP scheduling strategies, static and dynamic, and report the best performance. The static schedule assigns the same number of tasks to each thread, while the dynamic schedule does not assume the number of tasks per thread and assigns a task to a thread as soon as it is idle.
Our multi-threaded PBSM based on OpenMP supports arbitrary partition numbers and uses plane sweep to join the tiles. 
We adopt the one-dimensional PBSM, which partitions the data in one dimension and sweeps the data in the other dimension, because it has shown superior performance compared to the two-dimensional PBSM~\cite{tsitsigkos2019parallel}.

For GPU evaluations, we use cuSpatial~\cite{cuspatial}, the only open-source GPU-based spatial join library we found. cuSpatial only supports Point-in-Polygon tests (points join polygons using the within predicate) rather than joining two polygon datasets. Consequently, we restrict our GPU evaluations to these queries. Besides, cuSpatial employs quadtrees rather than R-trees as the index, because the boundaries of quadtree nodes can be computed simply with knowledge of the tree's levels and the node ID. However, this comes at the expense of an inferior indexing quality compared to R-trees~\cite{kothuri2002quadtree}. The quadtree index is only constructed on the point dataset, and the Point-in-Polygon queries are performed through the use of polygon batches serving as window queries.

\textbf{Hardware.}
The baseline CPU server contains an AMD EPYC 7313 16-core processor (7 nm) with a base frequency of 3 GHz and 256 GB of DDR4 memory. 
We run cuSpatial on an A100 SXM4 GPU (7nm) with 40 GB high-bandwidth memory (HBM).
We evaluated SwiftSpatial on an AMD Alveo U250 FPGA (16 nm) equipped with 64 GB of DDR4 memory (4 x 16 GB). 
Both the CPU server and the FPGA have comparable costs on AWS~\cite{aws_f1, aws_cpu}, at around 1.5 USD per hour on demand, while the GPU is more expensive, at around 4 USD per hour~\cite{awsp4}.
% \cite{aws_f1}
% \cite{aws_cpu}
% \cite{awsp4}

\subsection{End-to-end Spatial Join Performance}
% \vspace{-0.5em}

\begin{figure*}[t]
	% full width, can be adjusted
  \centering
  \begin{subfigure}[b]{\linewidth}
  \centering
    \includegraphics[width=1.0\linewidth]{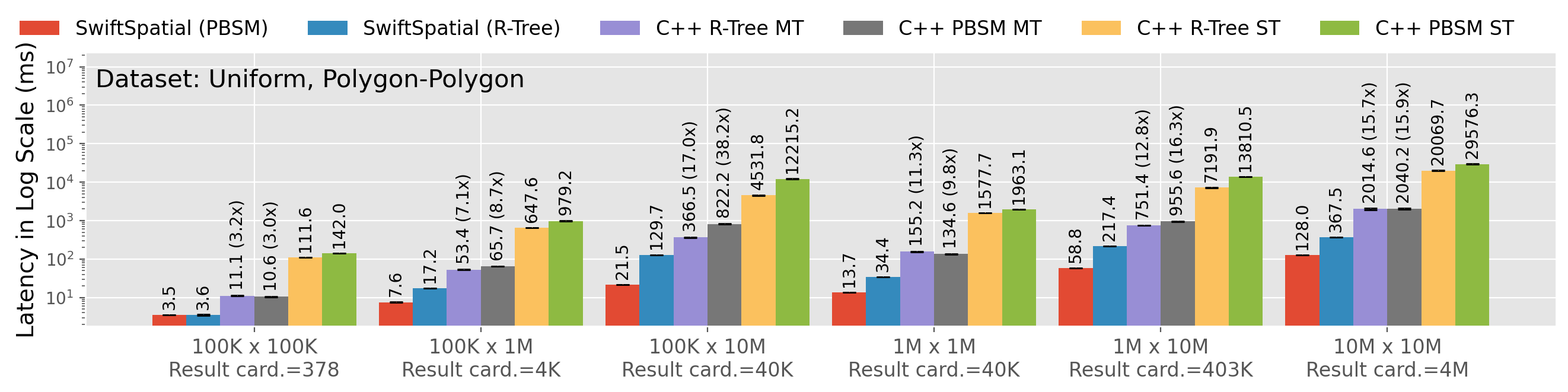}
    %\vspace{0.1em}
  \end{subfigure}
  \begin{subfigure}[b]{\linewidth}
  \centering
    \includegraphics[width=1.0\linewidth]{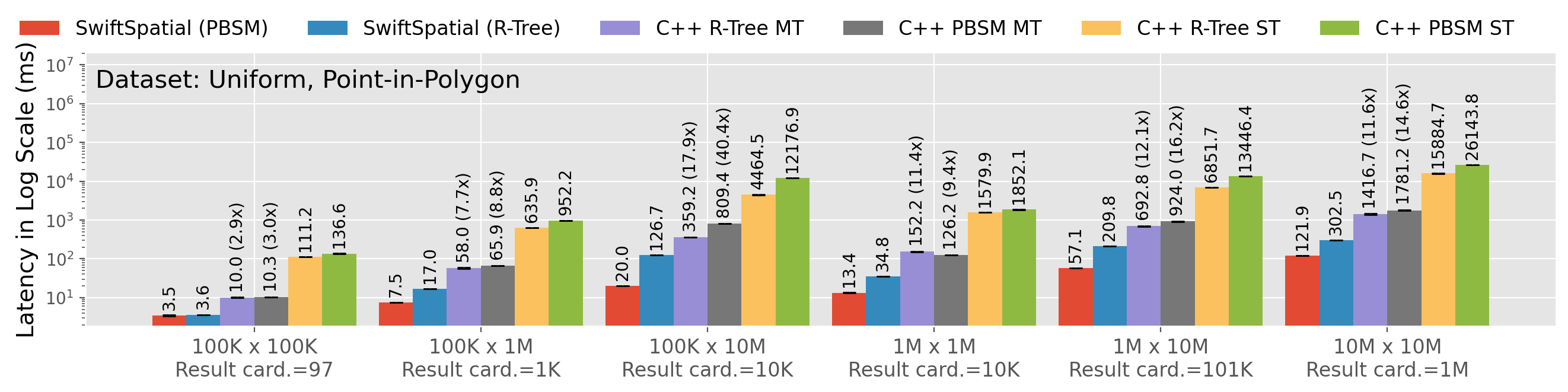}
    %\vspace{0.1em}
  \end{subfigure}
  \begin{subfigure}[b]{\linewidth}
  \centering
    \includegraphics[width=1.0\linewidth]{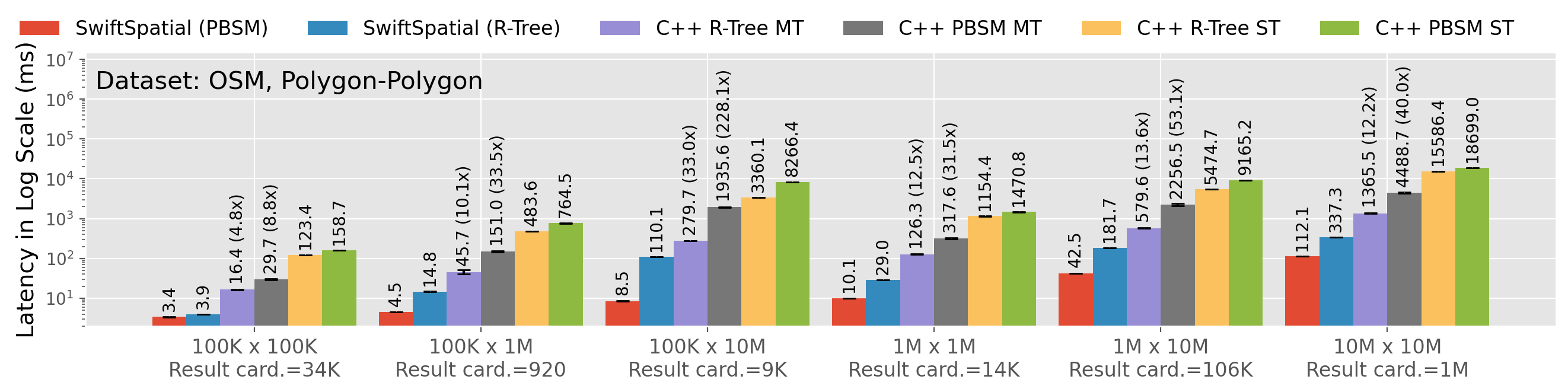}
    %\vspace{0.1em}
  \end{subfigure}
  \begin{subfigure}[b]{\linewidth}
  \centering
    \includegraphics[width=1.0\linewidth]{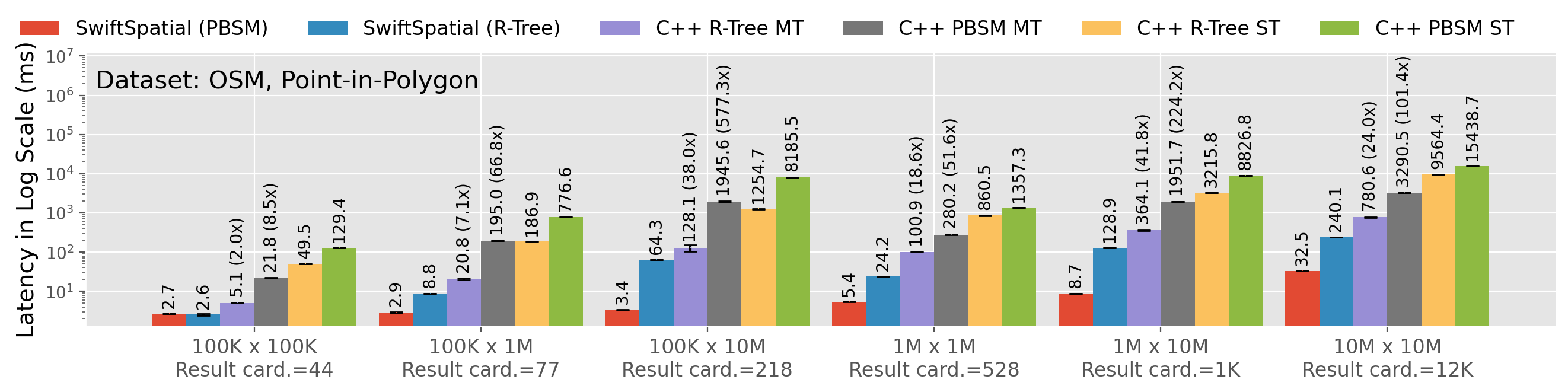}
    % %\vspace{0.1em}
  \end{subfigure}
  
  \vspace{-1em}
  \caption{Spatial join performance of SwiftSpatial compared to optimized CPU baselines across datasets, scales, and geometries.}
%   % \caption{The design of the SwiftSpatial accelerator primarily consists of a task scheduler, a set of identical join units, and various memory management units.} 
  \vspace{-1em}

  \label{fig:e2e-perf}
\end{figure*}

\begin{figure*}[t]
	% full width, can be adjusted
  \centering
  
  \begin{subfigure}[b]{0.7\linewidth}
    \includegraphics[width=\linewidth]{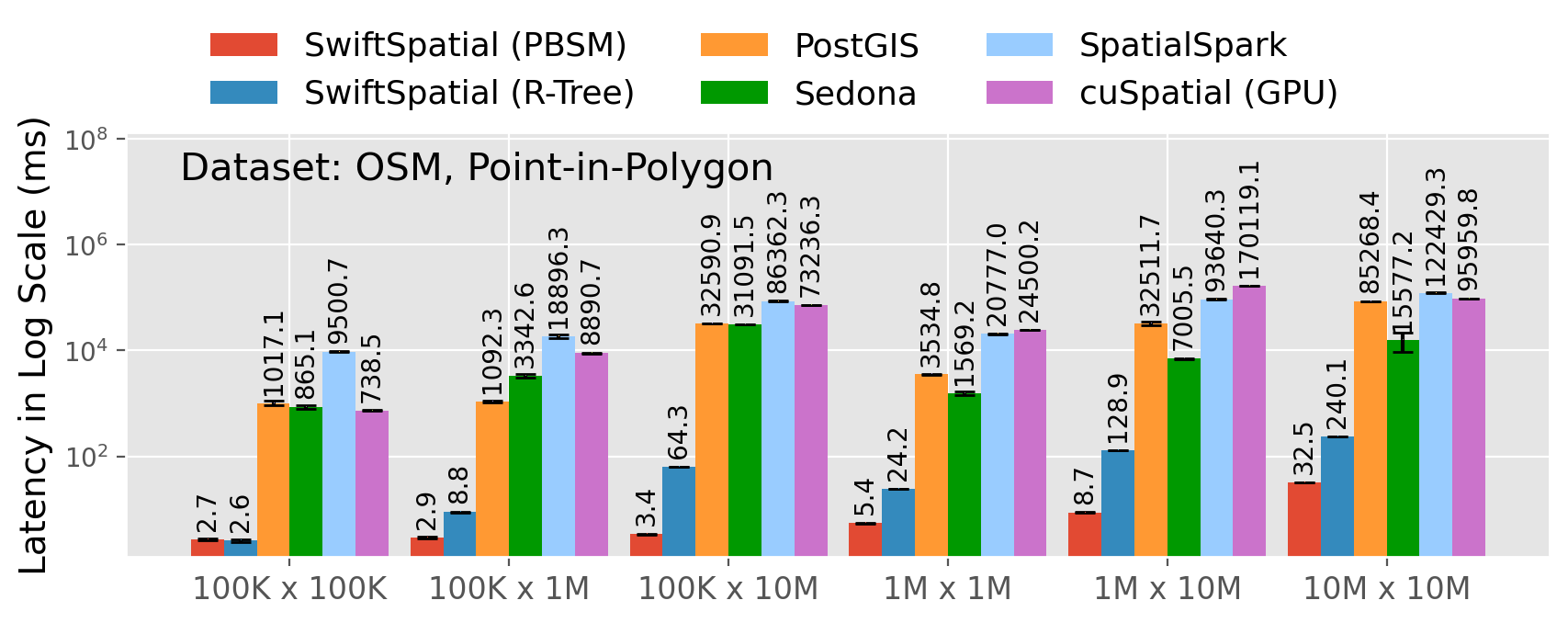}
    %\vspace{0.1em}
  \end{subfigure}
  
  \begin{subfigure}[b]{0.7\linewidth}
    \includegraphics[width=\linewidth]{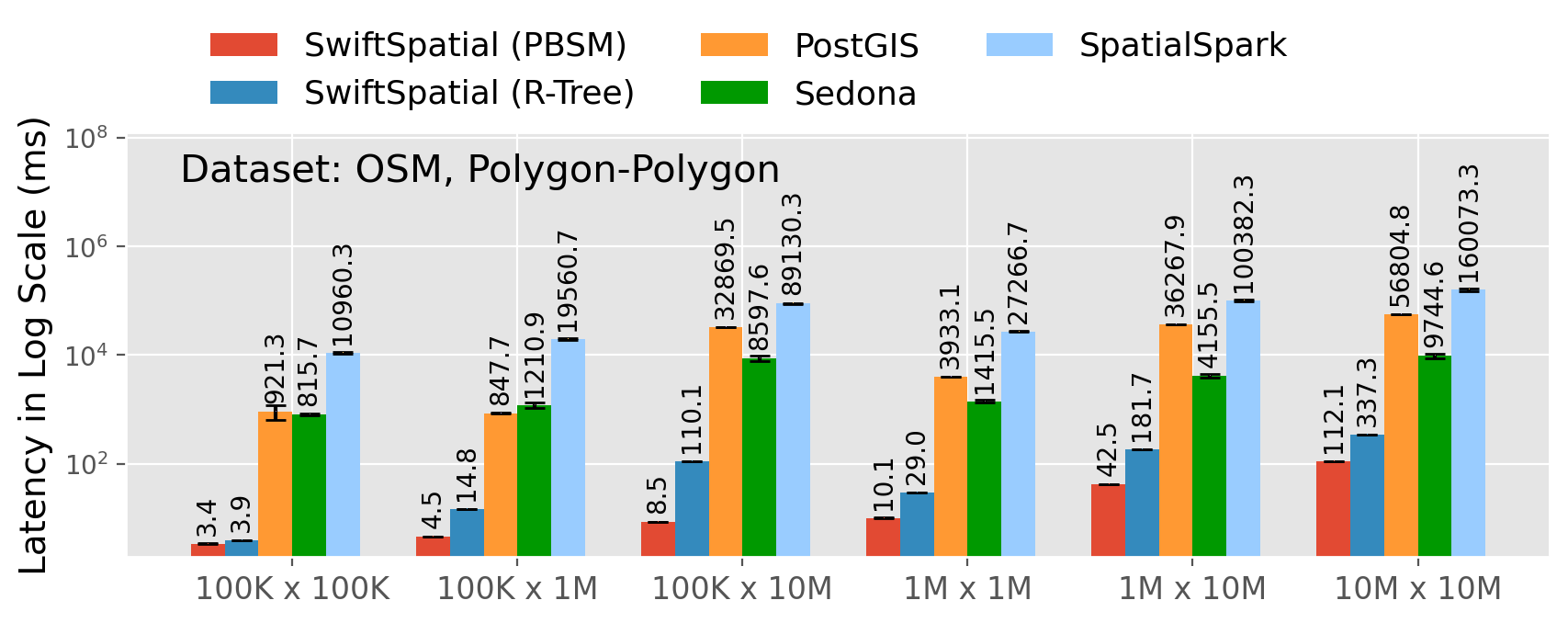}
    %\vspace{0.1em}
  \end{subfigure}
  
  \vspace{-1em}
  \caption{Spatial join performance of SwiftSpatial versus CPU- and GPU-based spatial data systems.}
  \vspace{-1em}
  \label{fig:e2e-perf-db}
\end{figure*}

We compare join latency between SwiftSpatial and the baseline CPU and GPU systems. We assume pre-constructed indexes on both datasets (R-trees and PBSM partitions), with associated costs reported separately in \S\ref{sec:index_cost}. For FPGA experiments, the join latency includes the join operation (kernel) and data movement (index and results) between the CPU and FPGA. For baselines, we measure only the join latency itself: the data and indexes are already loaded in memory (or GPU memory), and, for big data frameworks and PBSM, the data is already partitioned.

We report the best baseline and FPGA performance using the following configurations.
For C++ multi-threaded synchronous traversal, we tune parameters, including node sizes, traversal strategies, and OpenMP scheduling policies. After evaluating different traversal and scheduling strategies, we find that a combination of BFS and dynamic scheduling provides the best performance in the majority of our experiments. For both SwiftSpatial and the C++ synchronous traversal, the maximum R-tree node size (the number of entries a node can accommodate) is set to 16, which is shown to be optimal in a later section  (\S\ref{sec:node_size}).
For C++ multi-threaded PBSM, we evaluate different partition numbers ($10^2\sim10^5$) and partition-sweep directions (partition in one direction and sweep in the other) and report the best performance.
For SwiftSpatial PBSM, we set the maximum tile size to accommodate up to 16 objects (based on the geometric mean of the two input datasets) as this setting shows optimal performance.
With PostGIS, we configure the maximum parallel worker processes to 16, aligning with the available number of CPU cores on the server.
For SpatialSpark, we assess varying numbers of partitions and find that the optimal setting is 64.
In the case of cuSpatial, we fine-tune the average leaf node size in the quad-tree and set it as 128 for our experiments eventually. Since cuSpatial processes spatial join as window queries on the indexed point datasets, we use a batch size of 20K, such that the throughput is maximized without overflowing the GPU memory.

% ===== Speedup over all datasets =====
% Speedup over C++ multi-thread: 1.99 ~ 41.83 X
% Speedup over C++ PBSM multi-thread: 2.97 ~ 577.33 X
% Speedup over C++ single-thread: 19.37 ~ 395.93 X
% Speedup over C++ PBSM single-thread: 39.33 ~ 2428.92 X
% Speedup over PostGis: 188.24 ~ 9670.90 X
% Speedup over Sedona: 63.29 ~ 9225.98 X
% Speedup over SpatialSpark: 1095.46 ~ 25626.81 X
% Speedup over cuSpatial (GPU): 186.60 ~ 21731.85 X

% {\color{red} TODO}
\textbf{SwiftSpatial achieves significant speedup over all baselines.} 
Figure~\ref{fig:e2e-perf} compares SwiftSpatial against our optimized C++ spatial join implementation, while Figure~\ref{fig:e2e-perf-db} compares it with existing spatial data processing systems.
SwiftSpatial outperforms the multi-threaded synchronous traversal by 1.99$\sim$41.03$\times$, the multi-threaded PBSM by 2.97$\sim$577.33$\times$, the single-threaded C++ implementation by 19.37$\sim$395.93$\times$, the single-threaded PBSM by 39.33$\sim$2428.92$\times$, PostGIS by 188.24$\sim$9670.90$\times$, Apache Sedona by 63.29$\sim$9225.98$\times$, SpatialSpark by 1095.46$\sim$25626.81$\times$, and cuSpatial by 186.60$\sim$21731.85$\times$. 

As shown in Figure~\ref{fig:e2e-perf}, the PBSM version of SwiftSpatial shows the best performance across all datasets. 
PBSM allows SwiftSpatial to directly join tiles that are likely to yield results, unlike R-tree synchronous traversal, which involves multiple intermediate levels that require reading objects and writing intermediate results back.

% For CPU baselines, the multi-threaded synchronous traversal consistently performs the best or close to optimal among the baseline systems, except for a couple of cases (e.g., Uniform Point-in-Polygon 1Mx1M) which PBSM has a slight edge. 
For the CPU baselines, the multi-threaded synchronous traversal consistently delivers the best or near-optimal performance among the evaluated baseline systems.
While PBSM showcases near-linear scalability on Uniform datasets (14.5x latency reduction with 16 cores for 10Mx10M polygons), the single-threaded performance of PBSM lags behind that of synchronous traversal (same observation as in~\cite{sowell2013experimental}), thus PBSM, in the best case, can only match the performance of synchronous traversal with 16 cores. However, PBSM's scalability depends on data distributions: the scalability falters on OSM datasets with a skewed object distribution, even with large partition numbers.
Despite leveraging multi-core capabilities, PostGIS and the two big data frameworks tend to underperform single-threaded C++ synchronous traversal, as shown in Figure~\ref{fig:e2e-perf} and Figure~\ref{fig:e2e-perf-db}. This underperformance might be attributed to the overhead of abstractions and, for the big data frameworks, inefficiencies related to Scala compared to C++ and the data shuffling costs.

CuSpatial, despite being GPU-based, proves to be one of the slowest baselines, even underperforming the single-thread C++ implementations. For the GPU performance measurement, we already indexed the datasets and copied the data to the GPU memory prior to measuring the join performance. The low performance may stem from several factors. Firstly, the quadtree index implemented on the GPU is of lesser quality compared to R-trees~\cite{kothuri2002quadtree}. Secondly, cuSpatial only supports indexing for one dataset since it only supports point indexing. Consequently, polygons are processed as batch queries, resulting in less efficiency compared to synchronous traversal with both datasets indexed. Third, without knowing the number of join results in advance, cuSpatial resorts to over-allocating memory, limiting query batch sizes. For example, even though none of the datasets evaluated produce more than 300 million result object pairs (well below the 40 GB of GPU memory), the maximum batch size that passed all tests is only 20K. This limitation results in sub-optimal GPU core utilization.

% \vspace{-1em}
\subsection{Effect of Node Sizes}
\label{sec:node_size}
% \vspace{-0.5em}

We investigate the impact of varying R-tree node sizes, defined as the maximum number of objects or children a node can contain, on spatial join performance on CPU and FPGA.
Typically, smaller node sizes can result in more efficient search space pruning during synchronous traversal, thereby reducing the total number of predicate evaluations. However, this comes at the cost of increased random memory accesses to read more small nodes.
% For example, for polygon joins on the OSM 1M dataset, node sizes of 8, 16, and 32 lead to the reading of 524K, 244K, and 119K node pairs (decreasing random memory accesses), while evaluating 33M, 62M, 122M predicates between MBRs (increasing computation). 

\begin{figure*}[t]
	% full width, can be adjusted
  \centering

  \includegraphics[width=\linewidth]{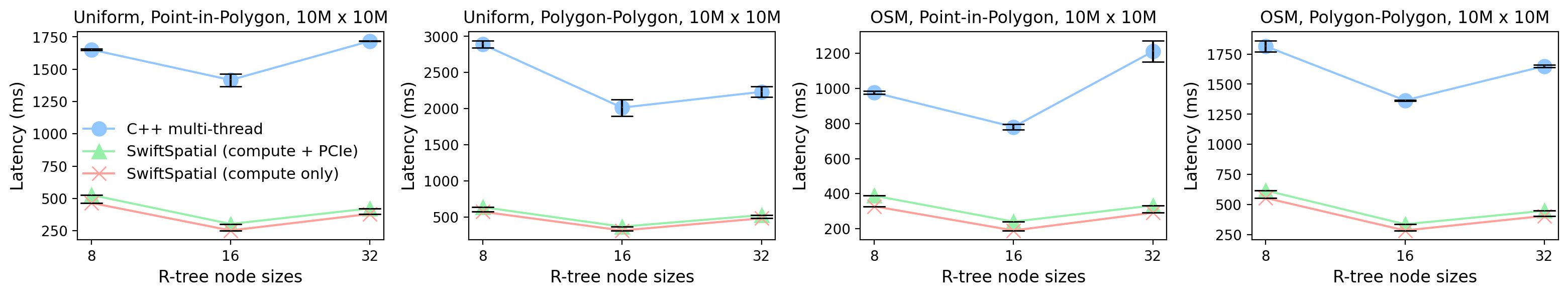}
  
  \vspace{-.5em}
  \caption{The effect of different R-tree node sizes on spatial join performance given 16 threads or 16 join units.}
  \vspace{-1em}
  \label{fig:perf_page_size}
\end{figure*}

Figure~\ref{fig:perf_page_size} shows the impact of node sizes on performance for both the 16-thread C++ synchronous traversal and the 16-join-unit accelerator. 
Both systems reach peak performance with a node size of 16. Despite smaller node sizes offering enhanced efficiency in pruning the search space, the increase in random memory accesses restricts the rate at which node pairs can be loaded into the processor or the accelerator. Conversely, with larger node sizes, the effectiveness of space pruning decreases, leading to more predicate evaluations between node pairs throughout the traversal.

% \vspace{-1em}
\subsection{Performance Scalability of the Join Units}
\label{sec:scalability}
% \vspace{-0.5em}

We evaluate the performance scalability of the join units in both synchronous traversal and PBSM by instantiating one to sixteen units on SwiftSpatial.
We first examine the impact of varying node sizes in relation to different numbers of join units. Subsequently, we evaluate the performance scalability across a range of node sizes in R-trees or tile sizes in PBSM.

\begin{figure*}[t]
	% full width, can be adjusted
  \centering
  \begin{subfigure}[b]{0.4\linewidth}
    \includegraphics[width=\linewidth]{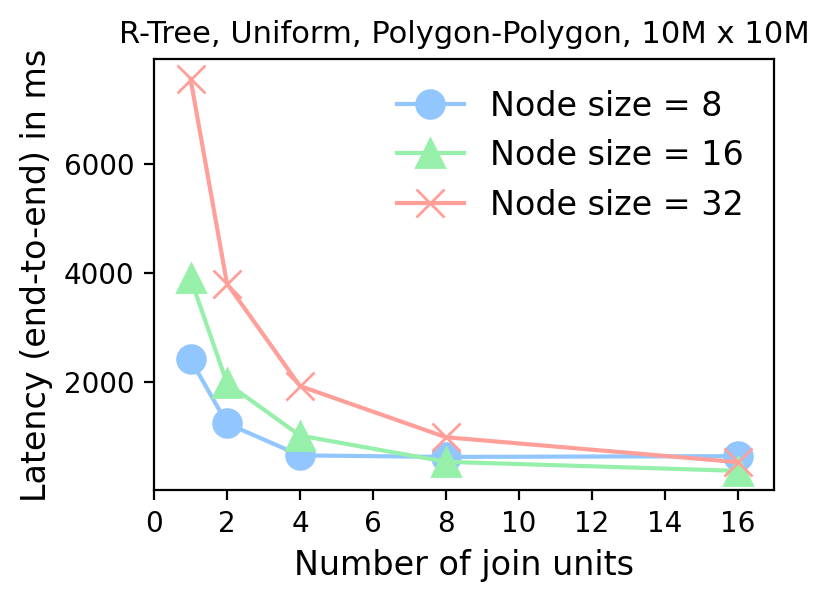}
    % \vspace{0.5cm}
  \end{subfigure}
  \begin{subfigure}[b]{0.4\linewidth}
    \includegraphics[width=\linewidth]{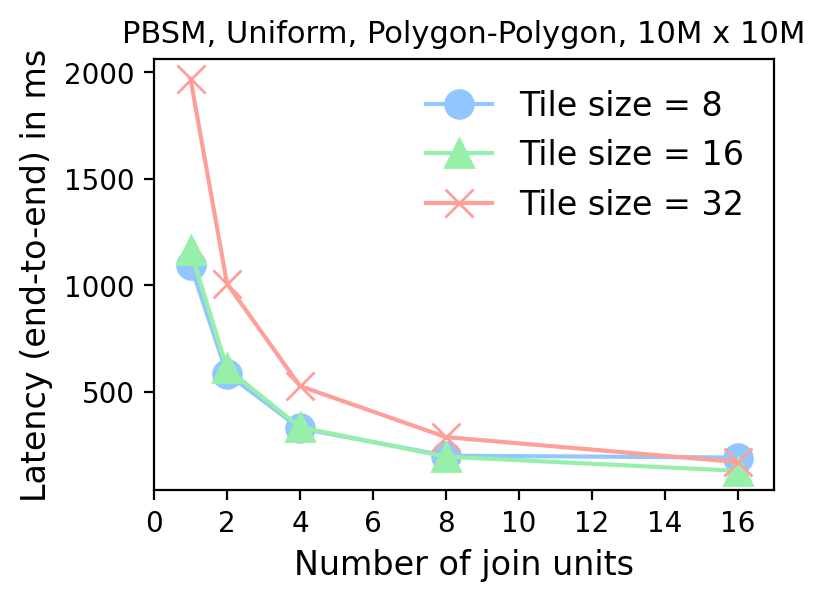}
    % \vspace{0.5cm}
  \end{subfigure}
  
  \begin{subfigure}[b]{0.4\linewidth}
    \includegraphics[width=\linewidth]{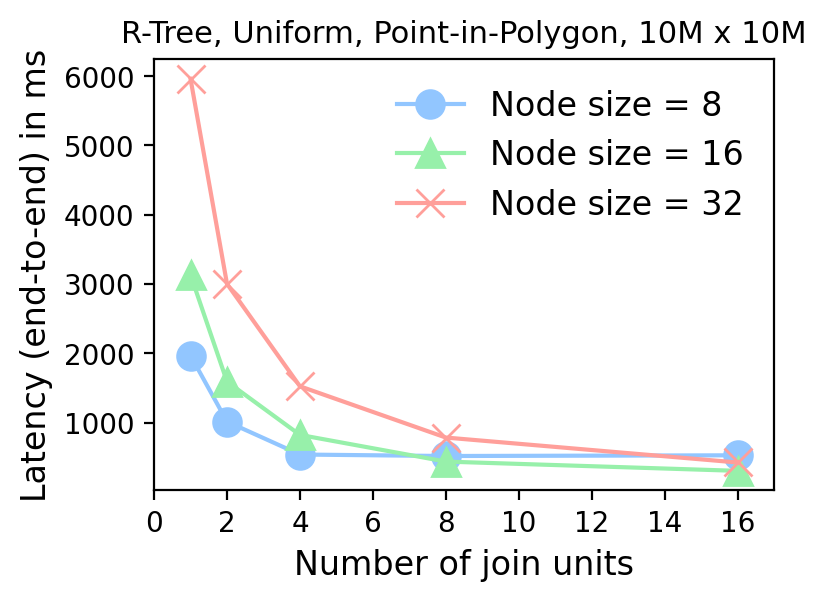}
    % \vspace{0.5cm}
  \end{subfigure}
  \begin{subfigure}[b]{0.4\linewidth}
    \includegraphics[width=\linewidth]{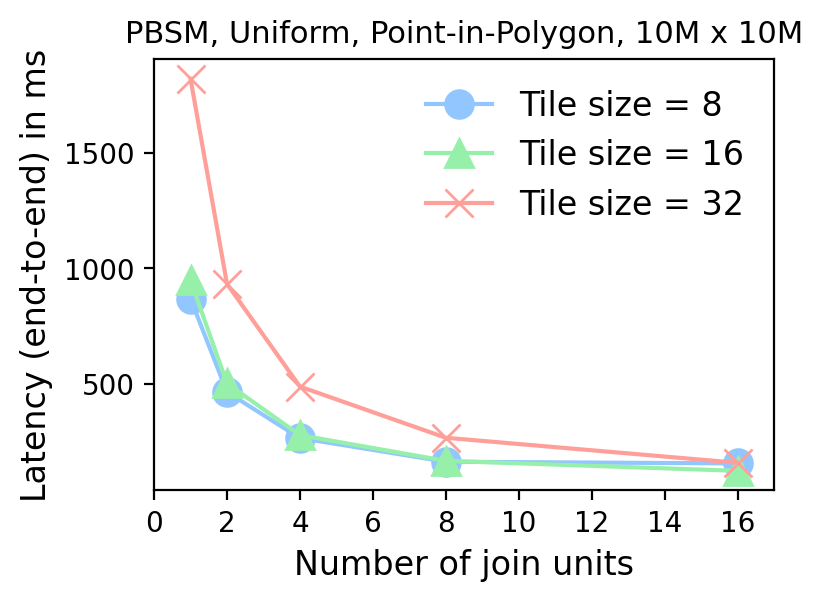}
    % \vspace{0.5cm}
  \end{subfigure}

  \vspace{-1em}
  \caption{The effect of R-tree node sizes or PBSM tile sizes when instantiating different numbers of join units.}
  \vspace{-1em}
  \label{fig:page_size_PE}
\end{figure*}

\textbf{The selection of optimal node sizes is related to the number of instantiated join units}. Figure~\ref{fig:page_size_PE} compares the end-to-end join performance using different datasets and node or tile sizes. For synchronous traversal for the Polygon-Polygon join on the Uniform dataset (upperleft of Figure~\ref{fig:page_size_PE}), the accelerator with a single join unit optimally favors a smaller node size of eight. As the quantity of join units escalates to eight, a node size of 16 emerges as the optimal selection. With an accelerator comprising 16 join units, a node size of 16 still retains its optimal status; however, the performance disparity between node sizes of 16 and 32 diminishes significantly compared to one join unit. %(305ms and 425ms, respectively). 
The rationale behind this lies in the fact that, when a lesser amount of join units are instantiated, the computation resources dedicated to predicate evaluation are restricted, thereby making R-trees with smaller nodes or PBSM with smaller tiles more desirable choices due to their potential to reduce computation effectively. Contrarily, when there is a substantial number of join units instantiated, these units end up competing for memory access. Consequently, spatial join with smaller nodes becomes memory-bound, leading to a situation where some units are idle, waiting for data.

\begin{figure}[t]
	% full width, can be adjusted
  \centering
  
  \begin{subfigure}[b]{0.38\linewidth}
    \includegraphics[width=\linewidth]{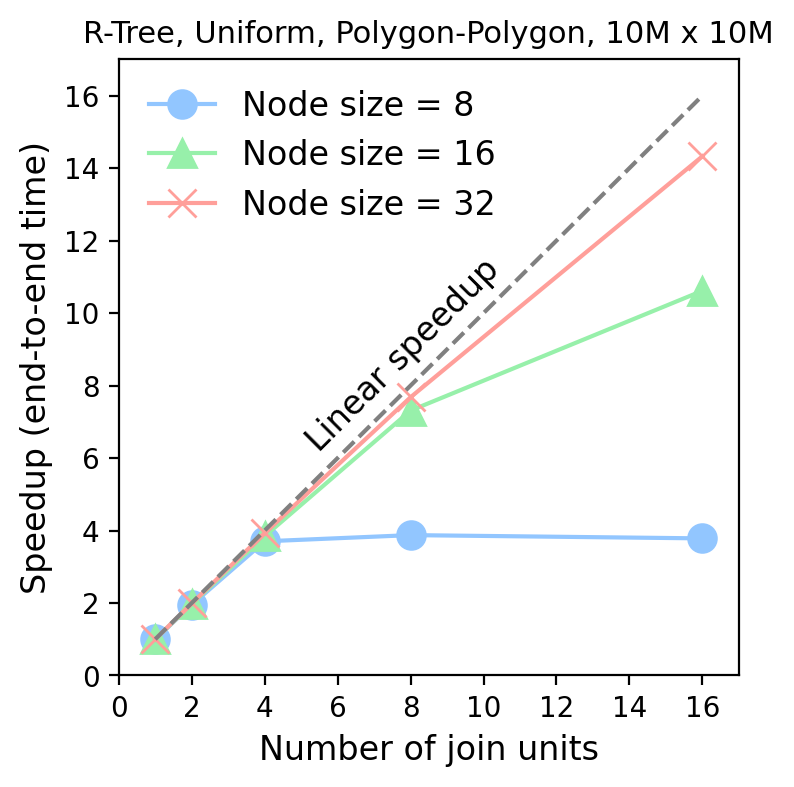}
    % \vspace{0.5cm}
  \end{subfigure}
  \begin{subfigure}[b]{0.37\linewidth}
    \includegraphics[width=\linewidth]{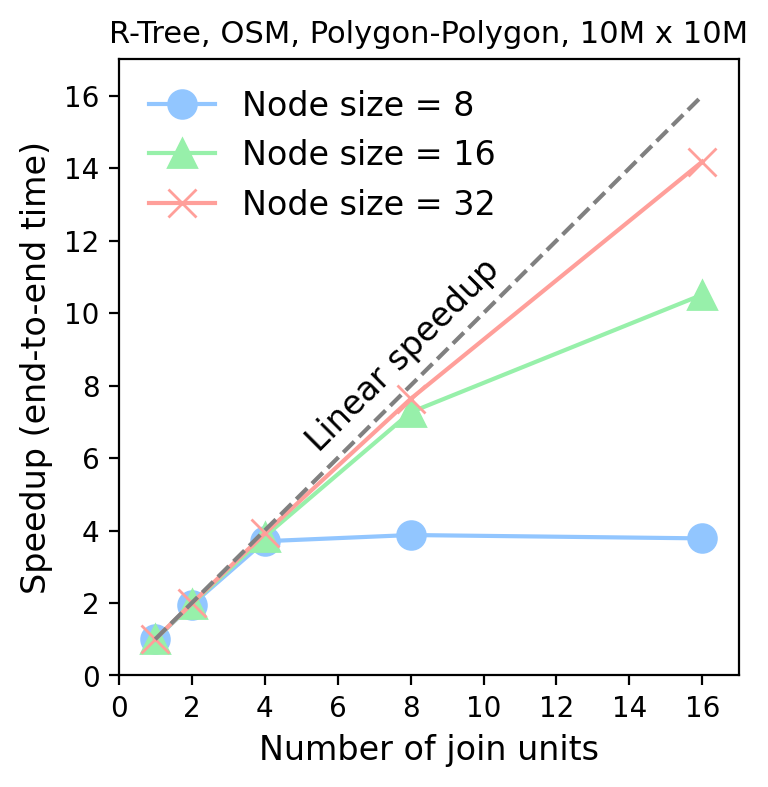}
    % \vspace{0.5cm}
  \end{subfigure}

  \begin{subfigure}[b]{0.38\linewidth}
    \includegraphics[width=\linewidth]{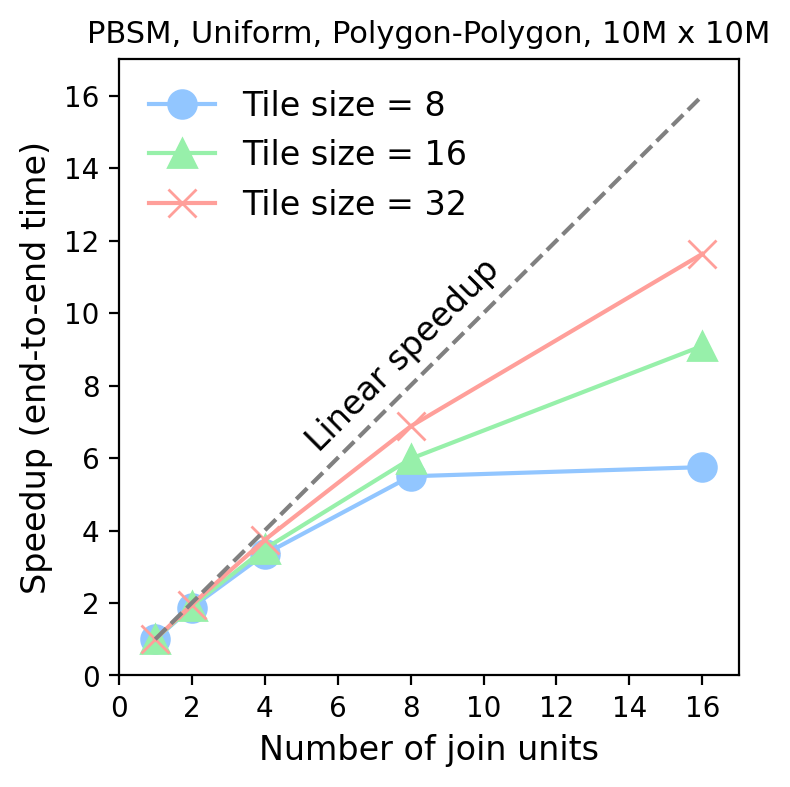}
    % \vspace{0.5cm}
  \end{subfigure}
  \begin{subfigure}[b]{0.37\linewidth}
    \includegraphics[width=\linewidth]{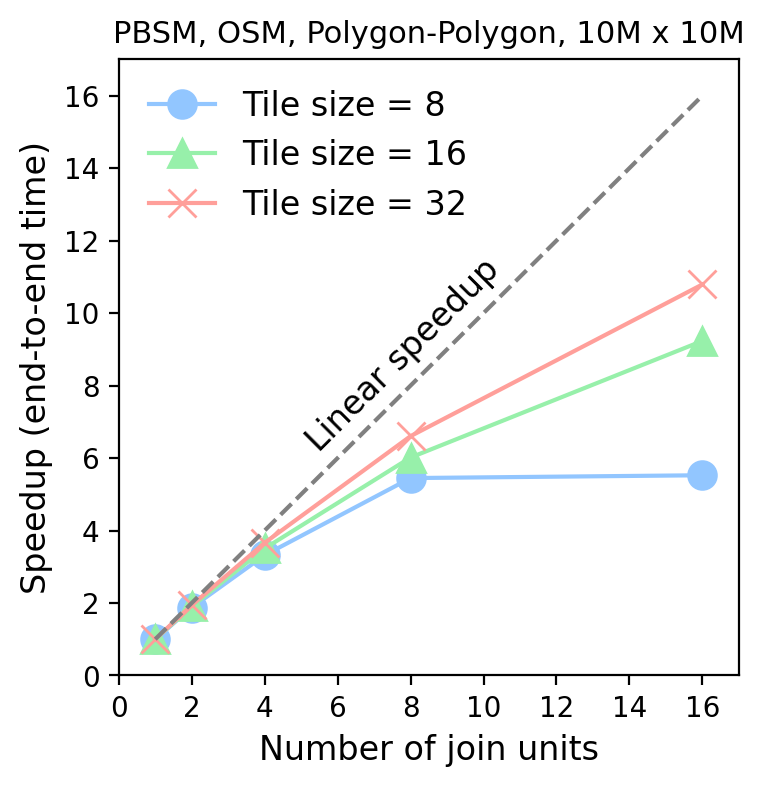}
    % \vspace{0.5cm}
  \end{subfigure}

  \vspace{-.5em}
  \caption{The performance scalability of the join units using different R-tree node sizes or PBSM tile sizes.}
  \vspace{-1em}
  \label{fig:scalability}
\end{figure}

\textbf{As node size increases, the performance scales better with the number of join units}.
Figure~\ref{fig:scalability} shows the speedup observed when instantiating multiple join units.
For synchronous traversal on the Uniform dataset (top left), the performance plateaus after instantiating four units when using a node size of eight, as the many short memory read operations cannot fully leverage the memory bandwidth. Conversely, with a node size of 32, the performance scales better by adding more units, demonstrating an almost linear progression up to 16 join units. 
The same performance trend has been observed on the OSM dataset (top right).
The PBSM implementation shows better performance scalability for small node sizes like eight (bottom of Figure~\ref{fig:scalability}), as it does not involve the intermediate result read and write operations in R-tree synchronous traversal.

\subsection{How Good is a Hardware Join Unit?}
\label{sec:join_unit_microbench}

In addition to the end-to-end performance evaluation, we microbenchmark the join units, specifically focusing on their efficiency in evaluating join predicates between nodes. For this purpose, we constructed an accelerator with a single join unit and fed R-tree node pairs of varying sizes into it.

\begin{figure}[t]
	% full width, can be adjusted
  \centering
  \begin{subfigure}[b]{0.35\linewidth}
    \includegraphics[width=\linewidth]{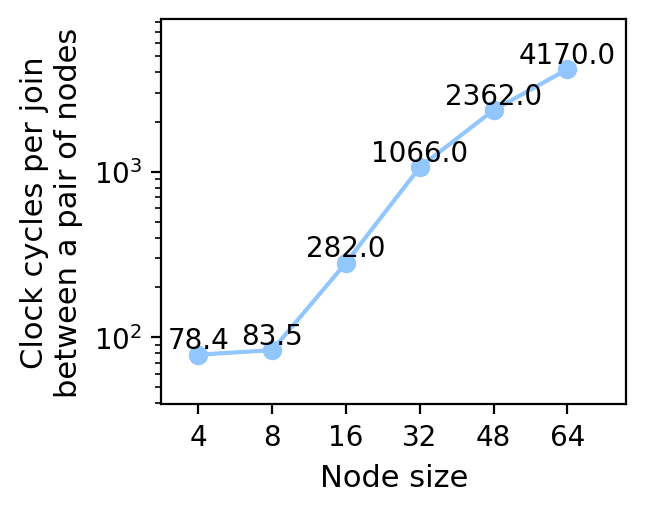}
    % \vspace{0.5cm}
  \end{subfigure}
  \begin{subfigure}[b]{0.33\linewidth}
    \includegraphics[width=\linewidth]{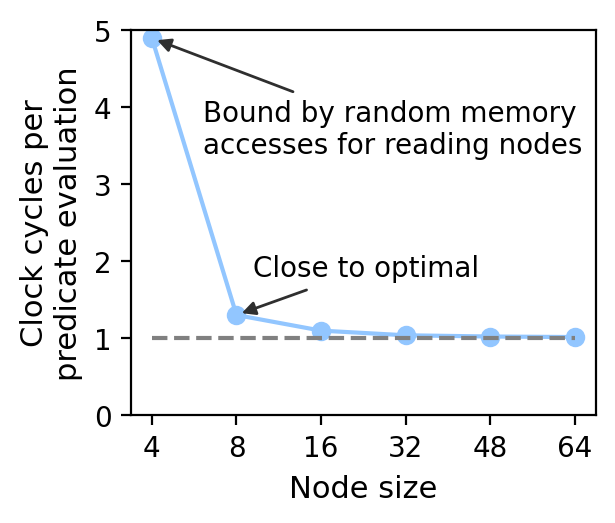}
    % \vspace{0.5cm}
  \end{subfigure}

  \vspace{-1em}
  \caption{Join unit can achieve near-optimal performance.}
  \vspace{-1em}
  % \caption{The design of the SwiftSpatial accelerator primarily consists of a task scheduler, a set of identical join units, and various memory management units.} 
  \label{fig:join_unit_microbench}
\end{figure}

\textbf{The join unit nearly achieves the ideal performance rate of one predicate evaluation per cycle.} Figure~\ref{fig:join_unit_microbench} shows the join unit performance given different node sizes. The plot on the left presents the number of clock cycles needed to perform a join between a pair of nodes, which includes the time needed for reading the data and evaluating the predicates. The plot on the right normalizes the performance as the average number of cycles needed per predicate evaluation. For instance, if both nodes contain 32 entries, then $32^2$ predicates are evaluated during the join operation. If this operation takes 1066 cycles to complete, then the number of cycles per predicate evaluation is $1066/32^2=1.04$. As shown in the plot on the right, the performance of joining small nodes (with four or fewer entries) is constrained by the random DRAM accesses to fetch the input node pairs. Conversely, the performance is close to the optimal value (one predicate evaluation per cycle) when a medium-sized node (eight or more entries) is employed. Here, the cycles per predicate range from 1.02$\sim$1.30 using node sizes of 8$\sim$64.

% Node sizes:  ['8', '16', '32', '64', '128']
% Time consumtion High / Low cardinality, CPU nested loop:  [1.5483871  1.472      1.47020531 1.48341885 1.52433669]
% Time consumtion High / Low cardinality, CPU plane sweep:  [1.29393939 1.4353289  1.74105561 2.45217991 3.35514171]
% low cardinality speedup: Plane sweep vs nested loop CPU:  [0.18787879 0.36954915 0.70740347 1.37704636 2.64134608]
% high cardinality speedup: Plane sweep vs nested loop CPU:  [0.22482436 0.37899073 0.59735504 0.83302881 1.20003895]
% FPGA speedup low cardinality / CPU nested loop low cardinality:  [2.9716685  3.54609929 3.7467167  3.83261391 3.87745803]
% FPGA speedup high cardinality / CPU nested loop high cardinality:  [4.60129316 5.21985816 5.50844278 5.6853717  5.91055156]
% FPGA speedup low cardinality / CPU plane sweep low cardinality:  [15.81694522  9.59574468  5.29643527  2.78321343  1.46798561]
% FPGA speedup high cardinality / CPU plane sweep high cardinality:  [20.46616851 13.77304965  9.22138837  6.82494005  4.92529976]
\begin{figure}%[t]
	% full width, can be adjusted
  \centering
  
    \includegraphics[width=0.55\linewidth]{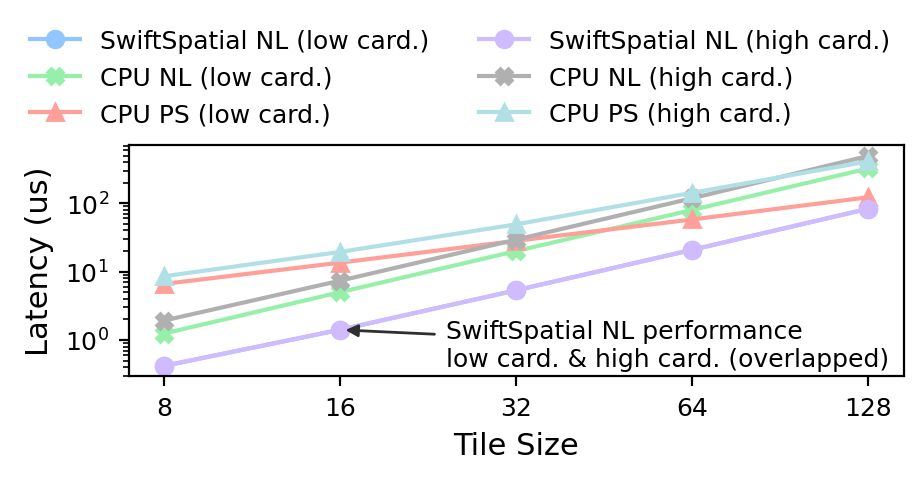}
    % \vspace{0.5cm}

  \vspace{-1em}
  \caption{Comparison of nested loop (NL) and plane sweep (PS) with different tile sizes and result cardinality.}
  \vspace{-1em}
  % \caption{The design of the SwiftSpatial accelerator primarily consists of a task scheduler, a set of identical join units, and various memory management units.} 
  \label{fig:tile_microbench}
\end{figure}

\textbf{The SwiftSpatial join unit exhibits a significant advantage over software-based nested loop join or plane sweep.} Figure~\ref{fig:tile_microbench} compares the latency of tile-level joins across varying tile sizes and result cardinalities, given a single SwiftSpatial join unit and a single-threaded C++ implementation. 
To modulate result cardinalities, we adjusted the edge lengths of the tiles and populated them with unit-length rectangles. As a reference, the low-cardinality configuration yields no results for tile sizes under 128, while the high-cardinality setup produces 2170 results when joining 128-object tiles.
Several observations can be made from Figure~\ref{fig:tile_microbench}. 
Firstly, even software nested loop join outperforms plane sweep for small and moderate tile sizes (less or equal to 32 and 64 for the low and high cardinality experiments, respectively). 
Secondly, the performance of plane sweep is more susceptible to data distributions, as the increasing number of objects in the sweep queues can substantially hinder its efficiency.
Thirdly, the SwiftSpatial join unit performance is consistent regardless of result cardinalities, as shown by the overlapped performance curve in Figure~\ref{fig:tile_microbench}.
Lastly, the fast predicate evaluation facilitated by the hardware join units leads to a significant performance advantage over software nested loop and plane sweep up to a moderate tile size (128), which is around an order of magnitude greater than typical R-tree node sizes.

% \vspace{-1em}
\subsection{Hardware Resource Consumption}
\label{sec:resource_consumption}
% \vspace{-0.5em}

Table~\ref{tab:resource} shows the hardware resource usage of SwiftSpatial, configured with different numbers of join units in the kernel. The kernel stands for user-space logic, while the shell refers to the FPGA infrastructure, including the memory controller, the PCIe controller, etc., consuming a constant amount of resources. The resource categories include Lookup Tables (LUTs) used for combinational logic, Flip-Flops (FFs) functioning as registers, Block RAMs (BRAMs) serving as on-chip memory, and Digital Signal Processors (DSPs) responsible for floating-point operations.

\textbf{On a data center FPGA such as U250, an accelerator kernel equipped with 16 join units consumes less than 30\% of the total hardware resources.} Notably, the most substantial resource consumption is Block RAMs (BRAMs), accounting for 28.05\% of the total resources, while other resources are far less utilized (below 4\%). 
This is attributed to the fact that each join unit in SwiftSpatial requires minimal resources (LUT, FF, and DSP), while the FIFOs interconnecting the join units, burst buffers, and read and write units all demand BRAM slices. These FIFOs are used to buffer data within the processing pipeline, thereby minimizing stalls.

\textbf{SwiftSpatial, when configured with a smaller number of join units, can be deployed on embedded FPGAs for edge spatial computing.} For example, PYNQ Z2, one of the lowest-end CPU-FPGA SoC on the market, contains 106,400 FFs, 53,200 LUTs, 140 BRAMs, and 110 DSPs. 
Even under a conservative assumption that 60\% of these resources can be allocated for user kernels, it would be feasible to host a SwiftSpatial kernel comprising one to two join units. By optimizing BRAM usage, such as employing shift registers in place of BRAMs to construct the FIFOs, it becomes possible to instantiate a SwiftSpatial kernel with up to four join units on the PYNQ Z2.

\begin{table}[t]
  \begin{center}
    \caption{SwiftSpatial consumes few FPGA resources.}
    \vspace{-1em}
    \label{tab:resource}
    \scalebox{0.9}{% scale down the table size to 80%
    \begin{tabular}{l r r r r r} % <-- Alignments: 1st column left, 2nd middle and 3rd right, with vertical lines in between
    \toprule
     & \textbf{LUT} & \textbf{FF} & \textbf{BRAM} & \textbf{DSP}\\
      \midrule
      % \hline % Should use \toprule, \midrule, and \bottomrule !
Kernel (1 PE)&	0.67\% &	0.44\% &	2.46\% &	0.16\% \\
Kernel (2 PE)&	0.87\% &	0.55\% &	3.65\% &	0.21\% \\
Kernel (4 PE)&	1.24\% &	0.75\% &	6.03\% &	0.34\% \\
Kernel (8 PE)&	1.96\% &	1.13\% &	10.79\% &	0.60\%\\
Kernel (16 PE)&	3.35\% &	1.60\% &	28.05\% &	1.12\% \\
% \midrule
Shell&	10.89\% &	9.21\% &	14.96\% &	0.11\% \\
% \midrule
Shell + Kernel (16 PE) &	14.24\%&	10.81\% &	43.01\% &	1.23\% \\

% Kernel (1 PE)&	11,587 (0.67\%)&	15,295 (0.44\%)&	66 (2.46\%)&	20 (0.16\%)\\
% Kernel (2 PE)&	15,011 (0.87\%)&	18,960 (0.55\%)&	98 (3.65\%)&	26 (0.21\%) \\
% Kernel (4 PE)&	21,505 (1.24\%)&	25,771 (0.75\%)&	162 (6.03\%)&	42 (0.34\%) \\
% Kernel (8 PE)&	33,860 (1.96\%)&	38,900 (1.13\%)&	290 (10.79\%)&	74 (0.60\%)\\
% Kernel (16 PE)&	57,842 (3.35\%)&	55,256 (1.60\%)&	754 (28.05\%)&	138 (1.12\%) \\
% \midrule
% Shell&	188,204 (10.89\%)&	318,385 (9.21\%)&	402 (14.96\%)&	13 (0.11\%) \\
% \midrule
% Kernel (16 PE) + Shell&	246,046 (14.24\%)&	373,641 (10.81\%)&	1,156 (43.01\%)&	151 (1.23\%)\\
\midrule
FPGA Total&	1,728,000 &	3,456,000&	2,688&	12,288 \\
\bottomrule
    \end{tabular}
    }
  \end{center}
	% \vspace*{-1.5em} % to shrink gap between figures
\end{table}

% \vspace{-1em}
\subsection{Power Consumption}
% \vspace{-0.5em}

We compare the power consumption between CPU, GPU, and FPGA for spatial join. For CPU, we measure the multi-threaded C++ implementation as it is the most performant software baseline; and we measure the GPU power consumption when running cuSpatial. In both cases, we executed the Point-Polygon join on the OSM 10M dataset continuously using a while loop and assessed the power using AMD RAPL (Running Average Power Limit) and NVIDIA System Management Interface, respectively. For the FPGA, we used Vivado to report the accelerator's power consumption.

\textbf{Despite its superior performance, the power consumption of SwiftSpatial is only 23.48W, 6.16$\times$ lower than the CPU and 4.04$\times$ lower than the GPU.} Thanks to the modest hardware resource utilization of SwiftSpatial (\S\ref{sec:resource_consumption}), the accelerator manages to sustain high performance while using minimal energy. The CPU power consumption (144.69W) is close to its Thermal Design Power (TDP) of 155W, indicating that the cores are fully utilized during the join operation. 
Contrarily, the GPU, despite having a 400W TDP, consumes only 95.01W during the join. This discrepancy likely stems from significant under-utilization of the many GPU cores: the substantial memory usage of cuSpatial limits the batch sizes during the join operation, thus restraining the core usage.

\subsection{Filtering and Refinement}

% Mode: pp
% Filter percentages: [36.262522097819684, 91.40155809073588, 77.01327487699918]
% Refine percentages: [63.73747790218033, 8.598441909264128, 22.986725123000824]
% --- Speedup: Dataset: OSM, pp Speedup: [1.40392968 6.29150228 3.41191299]
% Mode: pip
% Filter percentages: [93.10975609756098, 99.53284863637859, 98.62196765498653]
% Refine percentages: [6.890243902439025, 0.46715136362141, 1.3780323450134773]
% --- Speedup: Dataset: OSM, pip Speedup: [ 1.78844057 17.15406321 18.25057648]

While this study primarily focuses on the filtering phase of spatial joins, we also evaluate the refinement performance to demonstrate the end-to-end benefits of SwiftSpatial in a spatial join pipeline.

\begin{figure}[t]
	% full width, can be adjusted
  \centering
  \begin{subfigure}[b]{0.4\linewidth}
    \includegraphics[width=\linewidth]{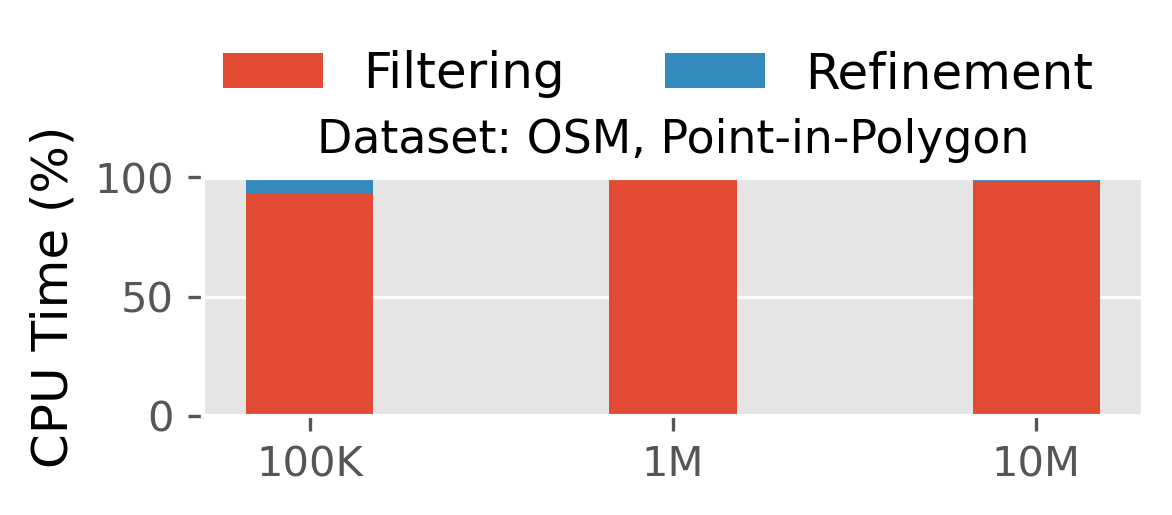}
    % \vspace{0.5cm}
  \end{subfigure}
  \begin{subfigure}[b]{0.4\linewidth}
    \includegraphics[width=\linewidth]{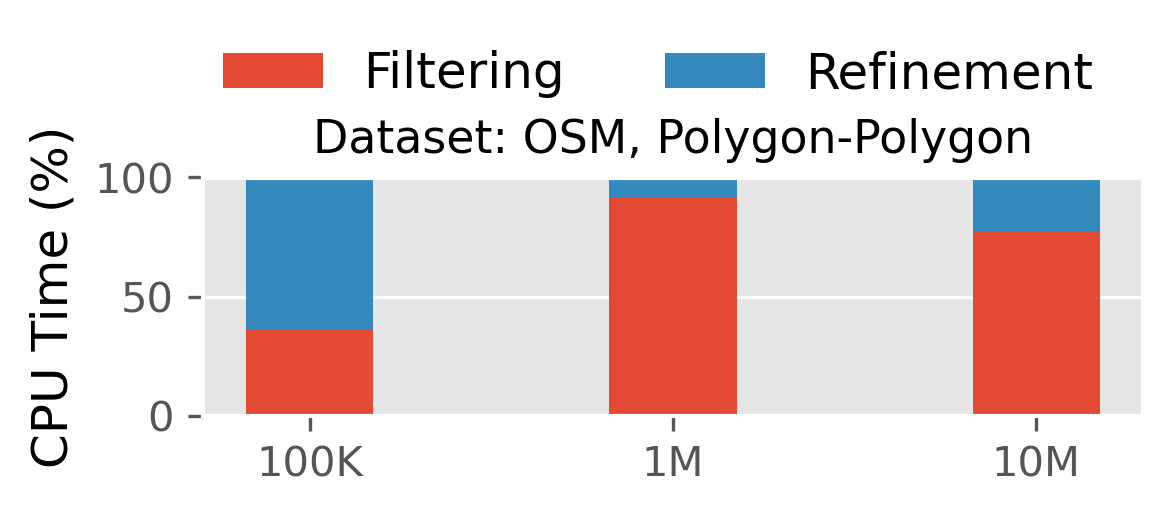}
    % \vspace{0.5cm}
  \end{subfigure}

  \vspace{-1em}
  \caption{Filtering versus refinement latency on CPUs.}
  \vspace{-1em}
  \label{fig:refinement_perc}
\end{figure}

Figure~\ref{fig:refinement_perc} shows the percentage of time spent on the refinement stage in an end-to-end spatial join on the OSM dataset using CPUs. Overall, the filtering phase often consumes more CPU time than the refinement phase for this dataset. However, the time distribution between these phases is strongly influenced by output cardinality. For instance, a 10M polygon-polygon join produces 1M result pairs after filtering, whereas a 10M point-in-polygon join yields only 12K result pairs. Consequently, the polygon-polygon join spends approximately 23.0\% of the total time on refinement, significantly more than 1.4\% for the point-in-polygon case.

\begin{figure}[t]
	% full width, can be adjusted
  \centering
  \begin{subfigure}[b]{0.4\linewidth}
    \includegraphics[width=\linewidth]{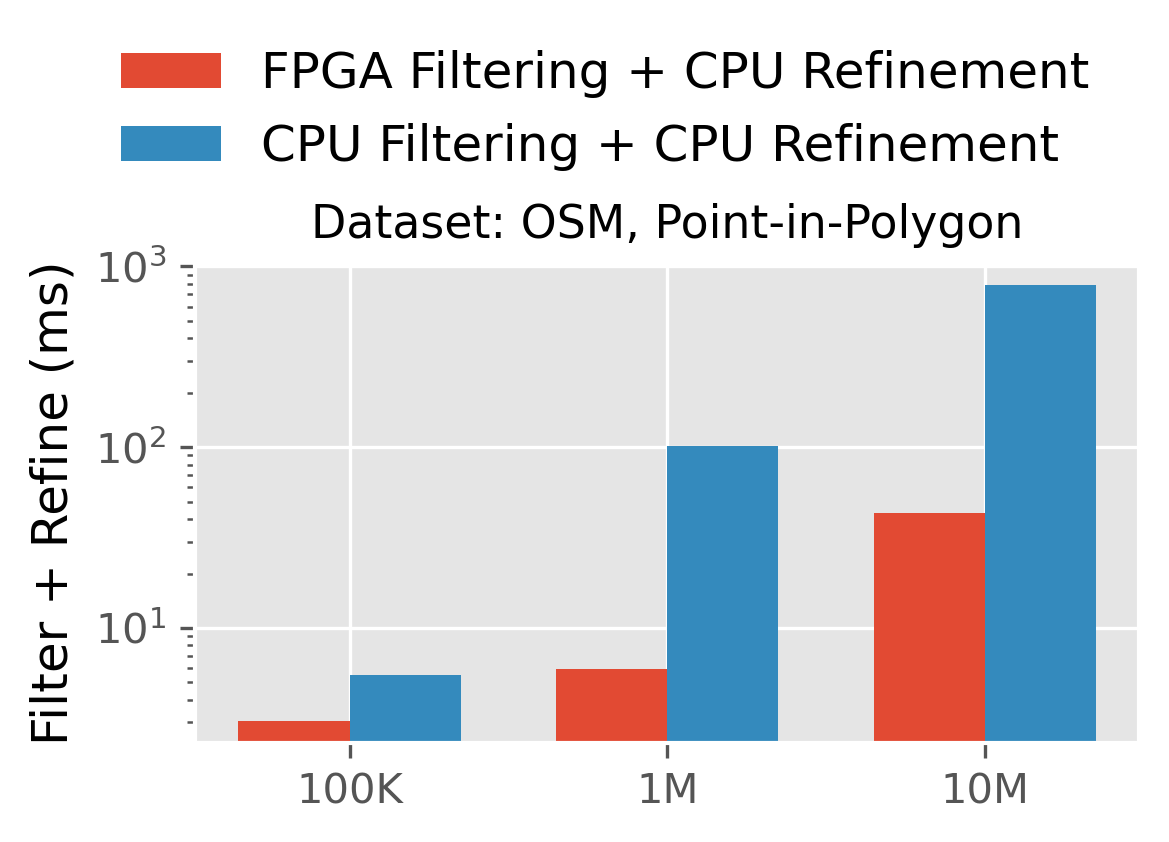}
    % \vspace{0.5cm}
  \end{subfigure}
  \begin{subfigure}[b]{0.4\linewidth}
    \includegraphics[width=\linewidth]{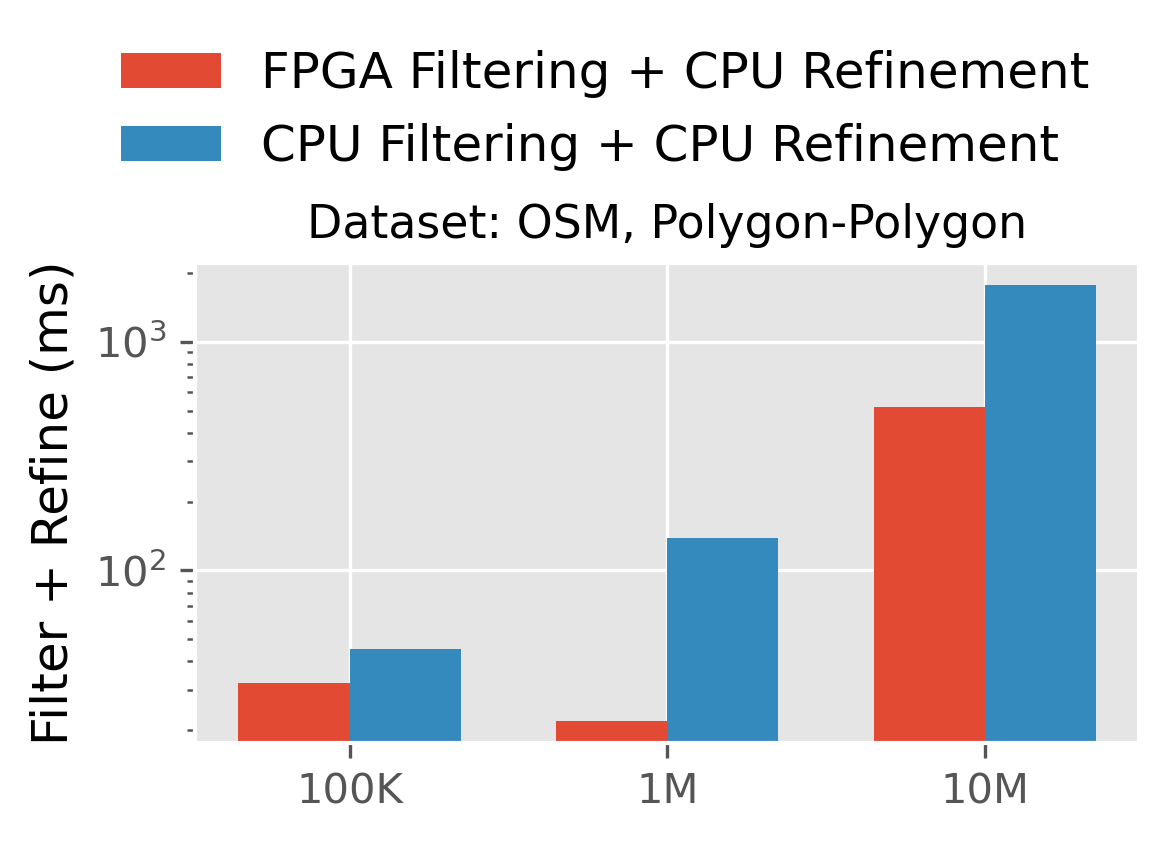}
    % \vspace{0.5cm}
  \end{subfigure}

  \vspace{-1em}
  \caption{End-to-end spatial join latency including filtering and refinement phases.}
  \vspace{-1em}
  \label{fig:e2e_filt_ref}
\end{figure}

Figure~\ref{fig:e2e_filt_ref} compares the performance of the end-to-end spatial join pipeline with and without SwiftSpatial, encompassing both filtering and refinement phases. In the case of SwiftSpatial, the filtered results are refined on a CPU server. Overall, SwiftSpatial accelerates the end-to-end spatial join pipeline by a factor of 1.4$\sim$18.3$\times$, depending on the filtering speedup and the fraction of time spent on filtering within the spatial join pipeline.

% \vspace{-1em}
\subsection{Index Construction Cost}
\label{sec:index_cost}
% \vspace{-0.5em}

% C++ parallel sort: https://www.modernescpp.com/index.php/multithreading-in-c-17-and-c-20

% {\color{red} TODO}

When interfaced with a spatial data management system, SwiftSpatial requires the R-trees to be constructed or the dataset to be partitioned just once by the CPU during the system's initialization phase. 
We study this one-time cost by implementing the R-tree index construction in C++ using the Sort-Tile-Recursive (STR) algorithm~\cite{leutenegger1997str}, which bulk loads the data to an R-tree using a recursive sorting and tiling strategy. We apply the parallel and vectorized sorting algorithm in the C++ Standard Template Library (STL). 
We also implemented a parallel data partitioning program: for CPU, it is a one-level partition, while for SwiftSpatial we add hierarchical partition to avoid tiles containing too many objects. 
Table~\ref{tab:index_cost} compares the time spent constructing indexes versus executing the join operation (using C++ multi-thread and SwiftSpatial) on several datasets of ten million objects, adopting a max node size or tile size of 16 as it demonstrates optimal join performance as evaluated in Section~\ref{sec:node_size} and~\ref{sec:scalability}. % on both CPUs and FPGAs.
Since the construction cost is typically higher than the join itself, spatial join based on R-trees would benefit more from iterative spatial joins~\cite{sowell2013experimental}, where at least one dataset is dynamic. In such scenarios, the R-tree is constructed once at the beginning, and subsequent insertions, updates, or deletions result in only minimal indexing overhead compared to bulk loading. 
For a one-off join, PBSM shows significantly less pre-processing overhead and thus would be a better choice.

\begin{table}[t]
  \begin{center}
    \caption{Time consumption of R-tree index constructions and data partitioning on ten-million-object datasets.}
    \vspace{-1em}
    \label{tab:index_cost}
    \scalebox{0.9}{% scale down the table size to 80%
    \begin{tabular}{l r r r } % <-- Alignments: 1st column left, 2nd middle and 3rd right, with vertical lines in between
    \toprule
     & R-tree & Hierarchical Partition & Partition \\
      \midrule
      % \hline % Should use \toprule, \midrule, and \bottomrule !
Uniform Point-Polygon   &	9,212 ms &	2,011 ms &	508 ms \\
Uniform Polygon-Polygon &	9,220 ms &	2,051 ms &	553 ms \\
OSM Point-Polygon       &	6,149 ms &	2,047 ms &	441 ms  \\
OSM Polygon-Polygon     &	6,903 ms &	2,662 ms &	456 ms \\
\bottomrule
    \end{tabular}
    }
  \end{center}
\end{table}

% \vspace{-1.5em}
\section{Discussion}
\label{sec:discussion}
% \vspace{-0.5em}

\textbf{Reusability and extensions for alternative join algorithms.} 
SwiftSpatial's modular design enables extensibility to other spatial join algorithms beyond PBSM and R-tree synchronous traversal. 
This versatility stems from the reusability of the accelerator's key component --- the hardware join units --- because joining small tiles is a major performance bottleneck in many spatial join algorithms~\cite{nobari2013touch, patel1996partition, tsitsigkos2019parallel, brinkhoff1993efficient}. 
To support alternative spatial join algorithms, only the on-chip scheduler for tile-level join dispatching needs to be updated. 
For instance, to implement algorithms with complex control flow, such as synchronous traversal (\S\ref{sec:scheduler}), the scheduler should manage (a) reading upcoming tile-level join tasks from memory, (b) dispatching these tasks to the instantiated join units, and (c) terminating the join process once no further tasks remain.

\textbf{Handling datasets larger than FPGA memory.}
So far, we have evaluated joins for which the indexed datasets and the join results fit in the FPGA memory. There are three potential solutions for larger-than-FPGA-memory joins.
The first solution involves data partitioning, akin to big data frameworks. The data is partitioned, and the join operation is segmented into several sub-tasks, which are handled by multiple FPGAs before the results are aggregated.
The second solution, given the significant speed advantage of FPGAs over software-based solutions, is that a single FPGA can process all data partitions iteratively.
The third solution is to scale up the FPGA memory capacity: the FPGA, as a computational device on its own, can be equipped with an extended memory capacity, even up to a terabyte of memory per device~\cite{cock2022enzian}. 

\textbf{Alternative spatial join algorithms on GPUs.}
While cuSpatial adopts quad-trees for spatial joins~\cite{cuspatial}, and several academic efforts have implemented R-tree-based spatial joins~\cite{prasad2015gpu, you2013parallel, kim2013parallel, luo2012parallel}, PBSM presents another viable option due to its embarrassingly parallel nature. 
PBSM has the potential to outperform both quad-trees and R-trees on GPUs, as it can be efficiently parallelized across partitions on GPUs using multiple streaming multiprocessors. 

However, we identify two performance challenges for GPU-based PBSM compared to SwiftSpatial.
The first challenge is memory management. Without prior knowledge of the output cardinality, GPUs would likely require a two-pass join process: one pass to determine memory allocation per partition and a second pass to write the results. This approach is necessary because each streaming multiprocessor needs to determine the start and end addresses for its write operations. In contrast, SwiftSpatial avoids this overhead with a single processing pass. Results from all join units are streamed into a single write unit, which dynamically manages memory access using a self-incrementing counter. As a result, join units in SwiftSpatial do not need to handle memory write addresses.
The second challenge is load balancing. Data skewness across PBSM partitions and the large number of streaming multiprocessors on GPUs can make it difficult to evenly distribute the workload. SwiftSpatial, however, mitigates load imbalance through its design: (a) it uses a smaller number of powerful join units, and (b) it supports both static and dynamic workload allocation policies. The dynamic policy ensures efficient load distribution by assigning new tile pairs to join units only when they become available.

\section{Related Work}

To the best of our knowledge, SwiftSpatial is the first specialized hardware architecture for spatial join. Since we have introduced the landscape of spatial join research in \S\ref{sec:background}, we now shift our focus to discussing related work in the realm of spatial data management techniques and hardware acceleration for data processing.

\textbf{Indexes for spatial data.}
The R-tree~\cite{guttman1984r}, introduced by Guttman in 1984, has been and continues to be one of the most popular index structures for managing spatial data. Various efforts have been undertaken to enhance the topology of the R-tree, aiming to speed up spatial queries. For instance, both STR~\cite{leutenegger1997str} and the Hilbert R-tree~\cite{kamel1993hilbert} improve R-tree topologies by more effective bulk loading strategies. On the other hand, the R\textsuperscript{+} tree~\cite{sellis1987r+} and the R\textsuperscript{*} tree~\cite{beckmann1990r} focus on optimizing subtree selection for insertion and node splits, and recent advancements have incorporated reinforcement learning to guide those policies~\cite{gu2023rlr}.
Apart from R-trees, there are other indexing techniques for spatial data. 
Quad-trees~\cite{samet1985storing, raulalexandrupersa2020adaptive} recursively divide space into four quadrants. 
KD-trees~\cite{orenstein1984class} split the space into two half-spaces along each of the k dimensions.

\textbf{FPGAs in data centers.} As Moore's Law has come to an end, substantial research has been devoted to processing data on hardware accelerators including FPGAs.
FPGAs have demonstrated potential not only in relational joins~\cite{chen2020fpga} but also in other database operations~\cite{jiang2023data} like regular expression matching queries~\cite{sidler2017accelerating}, decision tree traversal~\cite{owaida2019lowering}, sketching~\cite{kiefer2023optimistic, chiosa2021skt}, and data partitioning~\cite{kara2017fpga}. 
Beyond these core database operations, FPGAs can accelerate graph traversal~\cite{chen2021thundergp}, high-dimensional approximate nearest neighbor search~\cite{jiang2023co, jiang2023chameleon, jiang2024accelerating}, data encryption and compression~\cite{chiosa2022hardware}, embedding table lookups for recommender systems~\cite{jiang2021fleetrec, jiang2021microrec}, etc.
Beyond developing accelerators on FPGAs, researchers have also explored infrastructure on FPGAs such as virtualization~\cite{korolija2020abstractions, zha2020virtualizing, khawaja2018sharing} and utilizing FPGA as part of the software system infrastructure~\cite{guo2022clio, he2023accl+, wang2022fpganic, istvan2017caribou}.

% \vspace{-1.5em}
\section{Conclusion}
\label{sec:conclusion}
% \vspace{-0.5em}

We introduce SwiftSpatial, a hardware accelerator architecture for spatial join. 
It consists of specialized join units featuring hybrid parallelism, dedicated memory management units, and an on-chip scheduler that coordinates these components. 
Prototyped on FPGA, SwiftSpatial achieves a latency improvement of up to 41.03$\times$ compared to the best-performing software baseline, all while requiring 6.16$\times$ less power supply. The modular structure of SwiftSpatial enables (a) its flexible instantiation on both data-center-grade FPGAs and embedded systems and (b) its adaptability to various spatial join algorithms, indicating its extensive applicability in future spatial data management systems.

% \newpage

\begin{acks} 
We thank AMD for their generous donation of the Heterogeneous Accelerated Compute Clusters (HACC) at ETH Zurich (\url{https://systems.ethz.ch/research/data-processing-on-modern-hardware/hacc.html}), on which the experiments were conducted. 
% The HACC cluster is publically available to academic researchers: . 
\end{acks}

% \bibliographystyle{ACM-Reference-Format}
\bibliographystyle{plain}
\bibliography{ref}
% \bibliography{\jobname}

% \appendix


\end{document}